\let\oldvec\vec

\documentclass{raa}

\let\vec\oldvec

\usepackage{graphicx}
\usepackage{amssymb,amsmath}
\usepackage{txfonts}
\usepackage{natbib}
\usepackage{rotating}
\usepackage{color, colortbl}
\usepackage{sidecap}

\definecolor{LightCyan}{rgb}{0.88,1,1}

\definecolor{Gray}{gray}{0.9}

\begin{document}

   \title{A Review of Solar Type III Radio Bursts}

   \volnopage{Vol.0 (200x) No.0, 000--000}      
   \setcounter{page}{1}          

\author{H.A.S.~Reid\inst{1} \and H.~Ratcliffe\inst{2} }
\institute{SUPA School of Physics \& Astronomy, University of Glasgow, G12 8QQ, United Kingdom \and Centre for Space, Fusion and Astrophysics, Department of Physics, University of Warwick, CV4 7AL, United Kingdom}
  
   \date{Received~February~2014; accepted~March~2014}

\abstract{Solar type III radio bursts are an important diagnostic tool in the understanding of solar accelerated electron beams.  They are a signature of propagating beams of nonthermal electrons in the solar atmosphere and the solar system.  Consequently, they provide information on electron acceleration and transport, and the conditions of the background ambient plasma they travel through.  We review the observational properties of type III bursts with an emphasis on recent results and how each property can help identify attributes of electron beams and the ambient background plasma.  We also review some of the theoretical aspects of type III radio bursts and cover a number of numerical efforts that simulate electron beam transport through the solar corona and the heliosphere.}

\keywords{Sun: flares --- Sun: radio radiation --- Sun: X-rays, gamma rays --- Sun: particle emission}

\titlerunning{Solar Type III Radio Bursts}
\authorrunning{Reid and Ratcliffe}

\maketitle

\section{Introduction}


The Sun is the most efficient and prolific particle accelerator in our solar system. Electrons are regularly accelerated to near-relativistic energies by the unstable magnetic field of the solar atmosphere. Solar flares are the most violent example of such acceleration, with vast amounts of energy released when the solar magnetic field reconfigures to a lower energy state, releasing energy on the order of $10^{32}$~ergs and accelerating up to $10^{36}$ electrons per second in the solar atmosphere \citep[e.g.][]{Emslie_etal2012}.

\begin{SCfigure}
\centering
 \includegraphics[width=0.59\columnwidth]{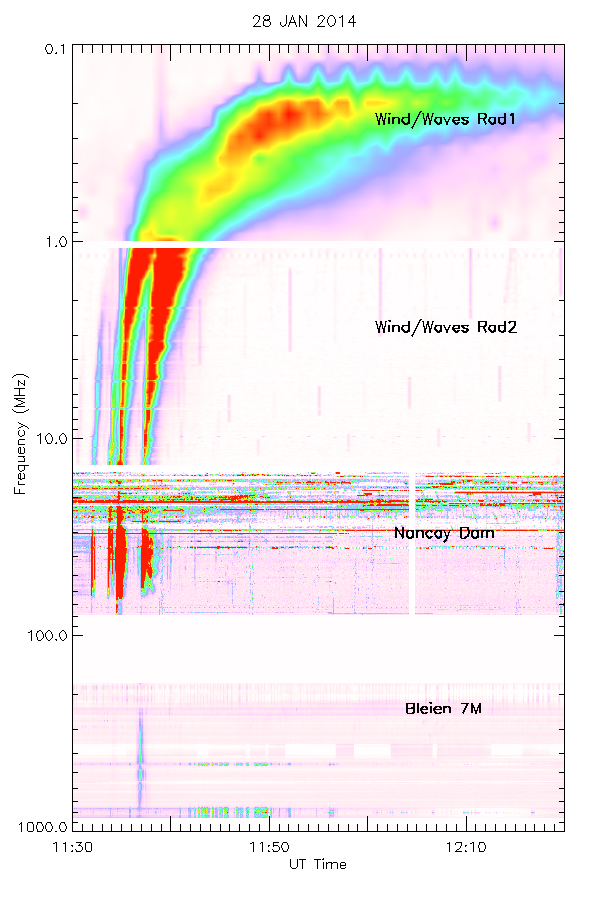}
\caption{An example of an interplanetary type III radio burst dynamic spectrum on the 28th January 2014.  The 900 - 200 MHz frequencies are observed by the Bleien telescope \citep{Benz_etal2009}.  The 80 to 15 MHz are obsered by the Nancay Decametre Array \citep{Lecacheux2000}.  The 14 to 0.1 MHz are obsreved by the WAVES experient onboard the WIND spacecraft \citep{Bougeret_etal1995}.  You can observe a number of different type III bursts starting at different frequencies from $<100$~MHz at 11:32 UT to from $>1$~GHz at 11:37 UT.  All the radio bursts merge into one at the lowest frequencies $< 1$~MHz.}
\label{fig:typeIII}
\end{SCfigure}

Solar radio bursts come in a variety of forms, classified by how their frequency changes in time, known as their frequency drift rate.  Initially three types of radio emission were named type I, II and III in order of ascending drift frequency \citep{WildMccready1950}, with types IV and V introduced later. Each type has subtypes that further describe the array of complex behaviour these radio bursts display.  In this study we are going to concentrate on the most prolific type of solar radio burst, the type III radio burst. These are a common signature of near-relativistic electrons streaming through the background plasma of the solar corona and interplanetary space, offering a means to remotely trace these electrons. Moreover, their dependence on the local plasma conditions means they act as a probe of the solar coronal plasma and the plasma of the solar wind.   The example type III burst in Figure \ref{fig:typeIII} shows their main features; they are very bright, transient bursts that usually drift from higher to lower frequencies over time

The precise mechanism for the acceleration of solar electrons is still debated but it is generally attributed to the reconfiguration of an unstable coronal magnetic field, resulting in the conversion of free magnetic energy to kinetic energy. This acceleration usually occurs at solar active regions but can occur when the magnetic field in a coronal hole interacts with the surrounding magnetic field (a process known as interchange reconnection).  Moreover, type III radio bursts are observed in relation with X-ray bright points \citep{Kundu_etal1994}.

\begin{figure}
  \centering
  \includegraphics[width=0.59\columnwidth]{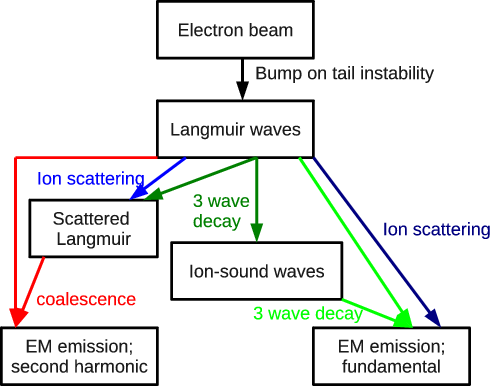}
  \caption{A flow diagram indicating the stages in plasma emission in an updated version on the original theory (adapted from \citep{Melrose2009})}
\label{fig:typeIII_flow}
\vspace{20pt}
\end{figure}

The first theory of type III bursts was described by \citet{GinzburgZhelezniakov1958}. They considered the two-stream instability of an electron beam which generates Langmuir (plasma) waves at the local plasma frequency that can be converted into electromagnetic emission.  They proposed that scattering by plasma ions would produce radiation at the plasma frequency (the fundamental component), while the coalescence of two Langmuir waves could produce the second harmonic. The theory has been subsequently discussed and refined by many authors \citep[e.g.][]{Sturrock1964,ZheleznyakovZaitsev1970, 1970SoPh...15..202S,1976SoPh...46..515S,1980SSRv...26....3M, 1983SoPh...89..403G,1985ARA&A..23..169D, 1987SoPh..111...89M}, but the basic two-step process, production of Langmuir waves followed by their conversion into EM emission, remains the same.  An overview of the dominant processes is shown in Figure \ref{fig:typeIII_flow}.

This \emph{plasma emission mechanism} is now the generally accepted model for type III burst generation. Analytical calculations over many years have shown it is certainly capable of explaining the drift, brightness and harmonic structure of bursts. Alternatives, such as strong turbulence effects like Langmuir wave collapse \citep[e.g.][]{1983SoPh...89..403G}; conversion of Langmuir waves directly into EM waves \citep[e.g.][]{Huang1998} and emission via the maser mechanism \citep[e.g.][]{2002ApJ...575.1094W} can explain some of the properties of type IIIs.  Where relevant, we assume plasma emission throughout this review.

The purpose of this study is to give an overview on the type III radio burst as a probe of both accelerated electrons and the background plasma it travels through.  In Section \ref{sec:observations} we give a detailed account of the observational characteristics of type III radio bursts and we describe how they relate to the properties of the generating electron beam.  Section \ref{sec:electrons} reviews in-situ plasma observations connected with type III bursts.  In Section \ref{sec:theory} we will cover some of the theoretical work that has been done on electron beams responsible for type III radio bursts.  In Section \ref{sec:em} we will give an overview of the theory involved with the generation of electromagnetic emission from plasma waves.  Finally we look towards the future in Section \ref{sec:conclusion} by covering the ``state of the art'' simulation codes and the next generation radio telescopes.

\section{Observed properties of type IIIs} \label{sec:observations}

A solar type III radio burst is a transient burst of radio emission that starts at higher frequencies and drifts to lower frequencies as a function of time.  Their durations, frequency extent and even how fast they drift varies from burst to burst.  Their size and intensity are different at different frequencies.  The nonlinear emission mechanism can lead to substructure in the dynamic spectra like clumpy emission and harmonic pairs.  There are a number of variant type III busts:  some have positive drift rates in frequency, some have different polarisation and some vary dramatically in frequency and time.  Other wavelengths of light can also be observed at the same time as type III bursts.

We review the main properties of the radio emission that we call a type III burst.  When relevant we also highlight what information can be deduced from the observations regarding both the exciting electron beam and the background heliospheric plasma.  Type III radio bursts have been studied intensely for the last 60 years so it is beyond our scope to mention all works that have been done on the subject.  We try to focus on the most recent works and to mention the range and variability of each property.

\subsection{Frequency and time} \label{sec:freqtime}

\subsubsection{Frequency extent}

The starting frequency of type III bursts varies dramatically from burst to burst.  During large solar flares type III bursts can start at frequencies in the GHz \citep[e.g.][]{1983ApJ...271..355B,StaehliBenz1987,Benz_etal1992,Melendez_etal1999}.  Typically type III bursts will start at 10s or 100s MHz and can start at even lower frequencies.  When two components are observed (see Section \ref{sec:harmStruc}), the fundamental often starts at lower frequencies than the harmonic, with the ratio of their onset frequencies often as large as $1:3$ or $1:5$ \citep{1977R&QE...20..989S}.   When type III bursts are observed at high frequencies they are usually in groups corresponding to multiple accelerated electron beams during solar flares.   The variation in starting frequency can be attributed to several causes.  Bursts at very high frequency (GHz range) are subject to absorption and have to be intense to be visible above the quiet-Sun background.  Moreover collisions in the background plasma damp both electrons and plasma waves making it harder to generate radio emission (see Section \ref{sec:el_sims}).  Favourable conditions for high frequency type III emission are acceleration regions at high densities that produce intense electron beams. High energy solar flares are the most common cause of such type III bursts.

The stopping frequency of type III bursts is equally as variable.  Some type III bursts will only exist at high frequencies above 100 MHz.  Such bursts are attributed to electron beams that are confined to the solar corona by a magnetic field with no access to the upper corona or the heliosphere (closed magnetic field).   Other type III bursts can make it to kHz frequencies, going down to 10s kHz or below, where 20 kHz corresponds to the plasma frequency near the Earth.  Such bursts are considered interplanetary type III bursts.  What frequency constitutes an interplanetary type III burst is not defined, as such a classification requires defining when the corona stops and interplanetary space starts.  Certainly if a type III burst makes it to frequencies below 1~MHz (roughly distance of 7 $R_\odot$) it should be considered ``interplanetary''.  Interplanetary type III bursts are usually made up of a number of distinct type III bursts at high frequencies that cannot be differentiated at lower frequencies (see Figure \ref{fig:typeIII}).  Using spacecraft, \citet{Leblanc_etal1995,Leblanc_etal1996} analysed the stopping frequency of type III radio bursts, finding that weaker radio bursts generally had higher stopping frequencies.  What process causes the stopping frequency is not clear but \citet{Leblanc_etal1995} hypothesise that is could be due to electron beam dilution and/or background density fluctuations.

\subsubsection{Burst Duration}

The duration of individual type III bursts varies inversely as a function of frequency.  Velocity dispersion of the electron beam exciter elongates the beam as a function of distance.  As such the beam spends more time at lower frequencies.  The rise and decay of type III radio emission in the interplanetary medium generally takes the form of a Gaussian total rise time $t_e$ followed by a power-law e-folding decay time $t_d$.  The general trend in emission is a shorter rise time $t_e < t_d$.  

A statistical study of rise and decay time between 2.8~MHz and 67~kHz was undertaken by \citet{Evans_etal1973}.  The study found with a least squares fit through the data the relations $t_e=4.0\times 10^8 f^{-1.08}$ and $t_d=2.0\times 10^8 f^{-1.09}$ where $t$ is in seconds and $f$ is in Hz.  A similar result was found by \citet{AlvarezHaddock1973b} using a number of different studies in the frequency range 200~MHz to 50~kHz.  They found a decay time $t_d=10^{7.7} f^{-0.95}$.  Rise and decay times that are roughly inversely proportional to frequency ($t_e\propto t_d\propto f^{-1}$) have been found at decimetric and microwave frequencies \citep{Benz_etal1983,StaehliBenz1987}.  It is important to note here that the rise and decay times are proportional to each other (longer rise times lead to longer decay times).  The power-law form of the decay time is currently unexplained as collisional damping of Langmuir waves would lead to a much longer decay time \citep[e.g.][]{Abrami_etal1990}.  There must be another process which accounts for either the spatial damping of Langmuir waves or the suppression of Langmuir waves inducing electromagnetic emission.  Harmonic emission is generally more diffuse in time, having a slower rise and decay time than fundamental emission for a given frequency \citep[e.g.][]{Caroubalos_etal1974}.  Given the differences in the emission mechanisms, particularly the potential involvement of rapidly damped ion-sound waves in fundamental emission, this difference is unsurprising.

\subsubsection{Frequency drift rate}

The defining property of a type III radio burst in comparison to other solar radio bursts is their high drift rate $df/dt$ (usually measured in ~MHz s$^{-1}$).  Exactly how the frequency drift rate is calculated is subjective with some authors using the drift rate of the type III burst onset and some authors using the drift rate of the type III burst peak flux.  One major survey of type III burst drift rate was done by \citet{AlvarezHaddock1973} who used the rise time of type III bursts between 3~MHz and 50 kHz and combined then with eight other studies up to 550~MHz.   They reported a least squares straight line fit to the frequency drift rate over all four orders of magnitude such that $df/dt=-0.01f^{1.84}$.  More recently a linear dependence ($df/dt\propto f$) was found for the frequency drift rate by \citet{Melnik_etal2011} between the frequencies 30~MHz to 10~MHz for a number of powerful radio burst occurring during a two month period in the solar maximum of 2002.  The value of the frequency drift rate was similar at 10 MHz to \citet{AlvarezHaddock1973} but smaller by a factor of two at 30~MHz.  Going to higher frequencies, studies by \citet{Aschwanden_etala1995} and \citet{Melendez_etal1999} looked at radio bursts between 100~MHz and 3000~MHz and found drift rates $df/dt=0.1 f^{1.4}$ and $df/dt= 0.09 f^{1.35}$ respectively.  Again, these values differ from \citet{AlvarezHaddock1973} but it must be noted here that \citet{AlvarezHaddock1973} did not use very many points greater than 200~MHz in their study.  Going to even higher frequencies a recent study by \citet{Ma_etal2012} looked at microwave type III bursts from 625~MHz to 7600~MHz, similar in range to events considered by \citet{StaehliBenz1987}.  \citet{Ma_etal2012} found a much faster frequency drift rate, $df/dt = -2.6\times 10^{-6}f^{2.7}$ in the range 635~MHz to 1500~MHz.  Both studies quote varying values from hundreds to 17000 MHz s$^{-1}$ at higher frequencies.

The drift rate of type III bursts can be used to assume the speed of the exiting electron beam.  Assuming a coronal (or interplanetary) density structure to obtain electron density as a function of distance $n_e(r)$, one can find frequency as a function of distance $f(r)$.  Assuming either fundamental or harmonic emission (see Section \ref{sec:harmStruc}) $df/dt$ can be converted to $dr/dt=v$ as a estimation of the speed of exciting electrons.  Derived exciter speeds for type III bursts are usually fractions of the speed of light and can range from $>0.5$~c in the corona \citep{Poquerusse1994,Klassen_etal2003} down to much slower velocities nearer the Earth.  Exciter speed derived by \citet{AlvarezHaddock1973} are fast, typically being $>0.2$~c.  Exciter speeds derived by \citet{Aschwanden_etala1995} and \citet{Melendez_etal1999} are in the regime of 0.14~c (7~keV).  These are slower than the exciter speeds of \citet{AlvarezHaddock1973}, however, \citet{Melendez_etal1999} specifically state using the maximum of type III emission and not the onset of type III emission.  The latter would lead to higher drift rates and might explain some of the discrepancy between the studies.  We expect the drift rate to be faster in the deep corona as the frequency of the background plasma is changing much faster with distance.

\subsection{Flux and size} \label{sec:fluxsize}

\subsubsection{Source flux}

At a given frequency the distribution of source flux of type III bursts varies over many orders of magnitude.  Statistically the distribution can be approximated as a power-law with a spectral index of 1.7 \citep{Nita_etal2002,Saint-Hilaire_etal2013}.  This spectral index has been found at GHz frequencies and frequencies at 100s MHz respectively.  It is the same spectral index that has been observed in X-ray flare energies \citep{Crosby_etal1993,Hannah_etal2008}.  This spectral index is also seen in self-organised criticality (SOC) studies \citep[e.g.][]{LuHamilton1991,Aschwanden2012}.

Statistically, the peak source flux of type III bursts increases as a function of decreasing frequency in the corona.  This increase goes up to roughly 1~MHz.  On average the peak flux then decreases as a function of decreasing frequency \citep{Weber1978,Dulk_etal1984,Dulk_etal1998}.  Exactly how the flux of individual type III bursts varies changes depending on the burst.  Some bursts have been observed to peak as low as 0.1~MHz while others peak as high as 5~MHz \citep{Dulk2000}.  An example spectrum can be found in \citet{Dulk_etal2001} showing a rise and decay of type III flux over two and a half orders of magnitude, starting around 50~MHz and ending at 20~kHz.  Recently \citet{Saint-Hilaire_etal2013} statistically examined type III burst source fluxes at 450~MHz to 150~MHz.  By fitting the power-law distributions of source flux they found the normalisation constants varied with frequency as $A\propto f^{-2.9}$.  The increase in source flux as a function of decreasing frequency can be attributed to a number of factors including the increasing ease to generate Langmuir waves, the increasing velocity dispersion of the electron beam as a function of distance and the decrease in collisions.

\subsubsection{Source Size} \label{sec:sourcesize}

The size of type III bursts increases with decreasing frequency.  Measurements at various frequencies for a variety of bursts give averages (half widths to 1/e brightness) of 2 arcmins at 432~MHz, 4.5 arcmins at 150~MHz \citep{Saint-Hilaire_etal2013}, 11 arcmin at 80~MHz, 20 arcmin at 43~MHz \citep{DulkSuzuki1980}, 5 degrees at 1~MHz,  50 degrees at 100~kHz \citep{Steinberg_etal1985} and 1~AU at 20~kHz (1~AU) \citep{Lin_etal1973}.  The electron beam exciter is guided by the solar/interplanetary magnetic field that expands as a function of distance from the Sun.  As the background electron density (frequency) decreases as a function of distance from the Sun, type III burst source sizes increase at lower frequencies.  The source sizes above 20~MHz are measured through ground based radio interferometers.  Radio emission below $10$~MHz does not typically make it through the ionosphere,  so source sizes here are measured using the modulation factor of antenna signal due to spacecraft spin.  

A comprehensive study of type III radio source sizes at low frequencies is undertaken by \citet{Steinberg_etal1985} who deduces an $f^{-1}$ variation of source angular size with observing frequency at frequencies below 2~MHz.  This is directly proportional to the distance from the Sun implying expansion in a fixed cone of $80^o$ with the apex in the active region.  However, there was a correlation between the local plasma frequency and the source size.  Scattering from density inhomogeneities in the background plasma expands the apparent source size and occurs within 0.2~AU at frequencies $<200$~kHz.  This results in electron beams expansion in a fixed cone of $30^o-40^o$ with the apex in the active region.  In the corona \citet{Saint-Hilaire_etal2013} also found an approximate $f^{-1}$ variation from the mean observed size between 150 and 432 MHz over 10 years of type III data using the Nan\c{c}ay Radioheliograph.  They also fitted the high end of the source size distribution with a power-law in both size and frequency, finding that the normalisation constant varied as $f^{-3.3}$.  It is not obvious why the exponent has such a high magnitude but it could be related to multiple sources merging into one at lower frequencies, given the constraint of 10 second data required for such a huge study of radio bursts.

\subsubsection{Brightness temperature}

The brightness temperature $T_b$ of a radio source is the temperature which would result in the given brightness if inserted into the Rayleigh-Jeans law.  The brightness temperature can be measured with knowledge of the flux density and the source size and is a commonly used metric in radio astronomy.  Type III bursts are characterised by their very large brightness temperatures that typically lie within the range $10^6$~K to $10^{12}$~K although it can rise as high as $10^{15}$~K \citep{SuzukiDulk1985}.  For 10 years of Nan\c{c}ay Radioheliograph data, \citet{Saint-Hilaire_etal2013} investigates the histogram of peak brightness temperatures within the range 150 - 450 MHz.  They find that the brightness temperature at all 6 frequencies varies as a power-law with spectral index around -1.8 (larger $T_b$ is less likely than smaller $T_b$).  The brightness temperatures varied from $10^6$~K to $10^{10}$~K with a few events $>10^{11}$~K.  At lower frequencies the trend is continued, with $T_b$ increasing with decreasing frequency up to around 1 MHz and then either decreasing or remaining constant \citep{Dulk_etal1984}.  The decrease of $T_b$ with decreasing frequency after 1~MHz cannot be explained due to increasing source size.  There is also a weak anti-correlation between rise times and $T_b$.  Fundamental type III emission is thought to produce higher $T_b$ than harmonic emission \citep{Dulk_etal1984,Melrose1989}, relating to both a larger flux density and a smaller source.

\subsection{Scattering, polarisation and harmonic structure}

\subsubsection{Scattering}

It is generally accepted that the source sizes of EM emission from the corona will be affected by scattering of radiation en route between source and observer. The rapid time variations in emission are inconsistent with a large source, and thus observations of higher frequency radio bursts with sizes of several arcminutes must be due to scattering.  Additional motivation comes from the observations summarised by \citet{1994ApJ...426..774B}, that no fine structure is observed in sources at scales below $20^{\prime\prime}$ at 1.5~GHz and around $40^{\prime\prime}$ at 300~MHz.  Ray-tracing analyses have been used to study the effect in context of type III emission \citep{1971A&A....10..362S,1972A&A....18..382S,1992A&A...253..521I} and a comparable analytical approach was developed by \citet{1999A&A...351.1165A}.  For example \citet{1972PASAu...2...98R} show that a point source at 80~MHz would be observed to be as large as 2 arcminutes, and the directivity of the radiation would be greatly reduced, particularly for fundamental emission, which is emitted into a relatively narrow cone.

A directivity study on 10 years of data from 1995 to 2005 was done by \citet{Bonnin_etal2008} using both the Ulysses and WIND spacecraft.  They examined about 1000 type III bursts observed in the frequency range of 940 kHz to 80 kHz.  They find that the radiation diagram axis shifts significantly to east from the local magnetic field direction, similar to previous observations \citep{Hoang_etal1997}, with this shift increasing for decreasing frequencies.  They attribute this to a density compression when the fast and slow solar wind meet, resulting in a transverse density gradient that bends radiation in the eastward direction.  They find no significant variation of directivity with solar activity and latitude.

\subsubsection{Harmonic Structure} \label{sec:harmStruc}

In a large number of observed radio bursts both the fundamental (F) and harmonic (H) components are seen. Bursts above around 100~MHz generally do not show two separate components and the observed component is thought to be the harmonic.  Below 100~MHz both fundamental and harmonic emission are frequently seen \citep[e.g.][]{Wild_etal1954a,Stewart1974, DulkSuzuki1980,SuzukiDulk1985, RobinsonCairns1994, RobinsonCairns1998}. Various theoretical estimates suggest that F emission is more common at large distances from the Sun, while the H component dominates closer, i.e. at higher frequencies \citep{RobinsonCairns1994,Dulk_etal1998}.  For emission at very high frequencies in the GHz and high MHz ranges, the absorption of radiation due to inverse bremsstrahlung becomes important.  Under the assumption of a locally exponential density profile the optical depth is easily calculated \citep[e.g.][]{1998SoPh..179..421K, 2014A&A_RatcliffeKontar}.  The F emission is highly suppressed above around 500~MHz, while the harmonic is only affected above 1~GHz or so \citep[e.g.][]{RobinsonBenz2000}.  In the 100-500 MHz range absorption is unimportant and cannot explain the rarity of F emission.   

The H-F ratio, naively expected to be 2:1, actually ranges from 1.6:1 to 2:1 with a mean near 1.8:1 \citep{Wild_etal1954a,Stewart1974}.  This may be explained by propagation effects altering the initial ratio of 2:1. Simply put, for EM radiation in plasma the group velocity tends to zero as the frequency approaches the plasma frequency whilst for frequencies much larger than $\omega_{pe}$ it tends to $c$. Thus the fundamental will be delayed relative to the harmonic emission from the same location \citep[e.g.][]{Dennis_etal1984}.  On the dynamic spectrum, the F component will be shifted to later times and the H-F ratio is correspondingly reduced.  The occasional instance of smaller ratios around 1.5:1 may suggest the two components are not F and H but rather the second and third harmonics \citep[see e.g.][]{1974SoPh...36..443Z,2012OAP....25..181B}.

\subsubsection{Polarisation} \label{sec:polarisation}

Type III bursts are usually weakly circularly polarized with H emission having a lower degree of polarisation than F emission \citep[e.g.][]{Mclean1971,SuzukiSheridan1977, DulkSuzuki1980,SuzukiDulk1985}.   \citet{DulkSuzuki1980} made a thorough analysis of polarisation characteristics of 997 bursts finding the average degree of polarisation of F-H pairs were 0.35 and 0.11 respectively while structureless bursts had only a polarization of 0.06.  The maximum F polarisation was around 0.6.  At high frequencies (164 - 432 MHz) polarisation has been found to peak either before or simultaneously with the peak flux but never after \citep{Mercier1990}.  Polarisation was also found by \citet{Mercier1990} to increase as a function of frequency, ranging from approximately $5~\%$ at 164 MHz to $15~\%$ at 432~MHz, in contrast to the results of \citet{DulkSuzuki1980}.  Type III bursts with nearly 100~$\%$ polarisation have been observed at microwave frequencies \citep{Wang_etal2003}.  For solar emission, any linear polarization in the source tends to be obliterated over any finite band of frequencies by differential Faraday rotation of the plane of polarisation during passage through the heliosphere \citep[e.g.][]{SuzukiDulk1985}.

Fundamental emission is generally produced very close to the plasma frequency and therefore below the X-mode cutoff.  It would thus have polarisation close to 1. Effects such as mode coupling due to magnetic fields \citep[e.g.][]{1964SvA.....7..485Z}, and scattering due to low-frequency waves \citep{1984SoPh...90..139W,1989SoPh..119..143M}, or kinetic Alfven waves \citep{2002A&A...390..725S} during propagation are certainly able to reduce the degree of polarisation but cannot explain why fully polarised emission is never seen. \citet{1984SoPh...90..139W} therefore proposed that the emission was depolarised to some extent within the source region itself, and that this was inherent to the emission process.   For harmonic emission, it was initially thought that the Langmuir waves involved would coalesce head on, and therefore \citet{1972AuJPh..25..387M, 1978A&A....66..315M} and the correction by \citet{1980PASAu...4...50M} concluded that harmonic emission must be weakly O-mode polarised. Later work by \citet{1997SoPh..171..393W} showed that relaxing the head-on condition allowed stronger polarisation, with less restriction on the participating Langmuir waves.

\subsection{Type III burst variants}

\subsubsection{Reverse and bi-directional type III bursts}

Ordinary type III bursts are produced by an electron beam travelling outwards from the Sun, along the magnetic field, and drift from high to low frequency. Electrons travelling downwards into the solar atmosphere instead encounter plasma of increasing density, and thus drift from lower to higher frequencies. These radio bursts usually have a high starting frequency $> 500$~MHz \citep{IslikerBenz1994,AschwandenBenz1997,RobinsonBenz2000} and fast drift rates due to the increased spatial gradient of the high density lower coronal plasma.   In some cases a reverse drift and a normal drift burst are produced at the same time, from the same acceleration region, leading to a bi-directional burst. These are relatively uncommon, with the study of \citet{Melendez_etal1999} finding only 5$\%$ of bursts which extended above 1~GHz had both components in contrast to $66\%$ which had a reverse slope component. 

The general suggestion to overcome strong inverse bremsstrahlung damping during escape is that the emission arises in dense loops embedded in less dense surrounding plasma \citep{Benz_etal1992}, which strongly increases the escaping fraction of radiation. Even in this case, it is likely to be only the harmonic component which may be observed \citep{Chen_etal2013b}.  Higher background densities increases the collisional rate and reduces the Langmuir wave growth rate.  An increasing background plasma density reduces the level of Langmuir waves generated by an electron beam \citep[e.g.][]{Kontar2001d}. High frequency type IIIs are thus less frequent and dimmer.  In the study of \citet{Melendez_etal1999} the peak fluxes were generally 10~sfu. 

Recent simulations were carried out by \citet{Li_etal2011} for bi-directional electron beams producing type III radiation.  As previously found from observations they find that downward going electron beams produce much less intense radio emission than upwards propagating electron beams.  That \citet{Melendez_etal1999} observes some instances of more intense emission in the reverse type III emission points towards an asymmetry for these events in the electron beam properties, like density.  \citet{Li_etal2011} also find the background plasma properties are hugely important to allow the emission at higher frequencies to escape the corona.

\subsubsection{Type IIIb bursts} \label{sec:IIIb}

Frequency fine structures are commonly observed in type III bursts, and generally referred to as a type IIIb burst. Rather than the smooth emission of a ``normal'' type III, the emission is fragmented and clumpy. In general, it is the F component which shows this structure, and fine structure in the H component in type IIIb bursts is very rarely observed \citep{DulkSuzuki1980}. Type IIIb bursts consist of a chain of emission bands, sometimes referred to as striae \citep{DelanoeBoischot1972}, each of which shows minimal frequency drift, whereas the chain as a whole drifts like a normal type III.  \citet{1975A&A....43..201D} found that the bandwidth of the striae in the 20-80~MHz range was commonly around 60-100~kHz, similar to their frequency separations. More recent observations by \citet{2010AIPC.1206..445M} show similar bandwidths for the striae of 50~kHz, but found strong variations in their duration, flux and frequency drift as a function of the source position on the solar disk.

The common belief \citep{SmithRiddle1975,Melrose21980,Melrose1983,Li_etal2012} is that density inhomogeneities in the background plasma create a clumpy distribution of Langmuir waves and are the cause of this fine structure. Harmonic emission, with its dependence on Langmuir wave backscattering, may be expected to show less modulation in this case. Fundamental emission, with its rapid growth and dependence on rapidly damped ion-sound waves, would be more susceptible. If density fluctuations are the cause then we may infer this is less intense closer to the Sun, leading to the absence of type IIIb's at higher frequencies.

Fine structure in the form of ``sub-bursts'', also in the decimeter range, was reported by e.g. \citet{2005A&AT...24..391M,2007HiA....14..365M}. In this case, the main burst envelope shows weak emission, and superposed on this are more intense stripes, with rapid frequency drifts on the order of 1 MHz s$^{-1}$, comparable to the main burst drift rate of around 3 MHz s$^{-1}$. These cannot be ascribed to density turbulence, and are as yet unexplained.

\subsubsection{Type U and J bursts}

\begin{SCfigure}
  \centering
  \caption{An example of a type J burst from the Glasgow Callisto spectrograph on the 13th March 2013.  Note the reverse in the frequency drift rate.}
  \includegraphics[width=0.59\columnwidth]{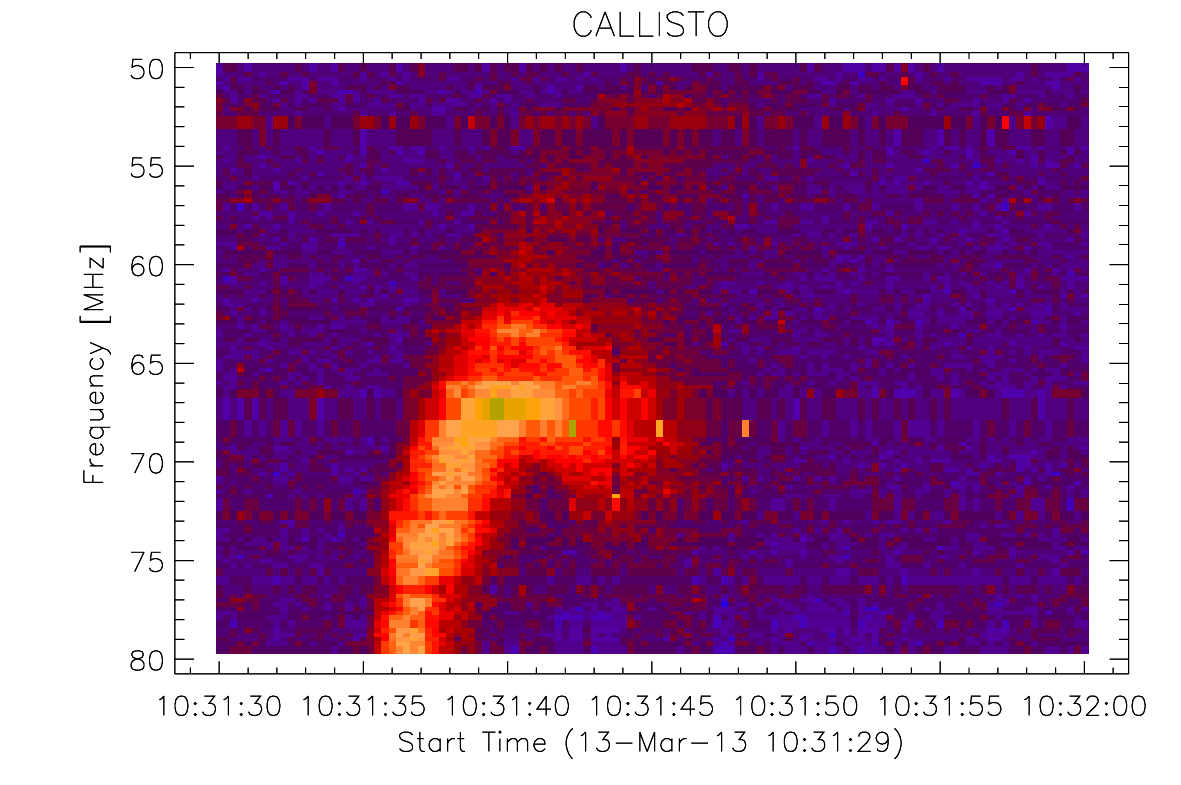}
\label{fig:type_u}
\end{SCfigure}

The frequency drift rate of radio bursts has been observed to change sign during a normal type III burst, taking the shape of an inverted U or J \citep{MaxwellSwarup1958} (Figure \ref{fig:type_u}).  These bursts are believed to be electron streams travelling along magnetic fields confined to the corona \citep[e.g.][]{KleinAurass1993,Karlicky_etal1996}.  For the J bursts, the radio emission stops when the electron beam reverses direction but with U bursts it continues to higher frequencies.  The rate of occurrence is very low and they generally occur in harmonic emission within the range 20-300~MHz, although fundamental emission has been observed \citep[e.g.][and references therein]{LabrumStewart1970,AurassKlein1997}.  The rising and descending branches have spatially separated sources \citep{AurassKlein1997}.    Their polarization is usually below 10~$\%$, agreeing with the properties of harmonic emission.  Similar to reverse drift bursts their low occurrence could be to do with increased difficulties to generate Langmuir waves in an increasing density gradient.

\subsubsection{Type V bursts}

Closely related to the type III burst are type V bursts, classified due to their long durations (minutes) and wide spectra \citep{Wild_etal1959}.  The type V emission appears as a continuation of a type III burst in the dynamic spectra.  Type V bursts are important because their explanation has to be consistent with any model of type III bursts.  Type V bursts appear at low frequencies below 120~MHz and generally have 1-3 minute durations \citep{Dulk_etal1980}.  The size of type V bursts increases rapidly with decreasing frequency, with full width at 1/e brightness on average 105 arcmins$^2$ at 80~MHz and 300 arcmins$^2$ at 43~MHz \citep{Robinson1977} similar to type III bursts \citep{DulkSuzuki1980}.  Type V bursts have also been observed to move relative to the disk surface at speeds $\approx 2~\rm{Mm~s}^{-1}$ \citep{WeissStewart1965}.  A similar problem to type IIIs occurs for the decay of type V emission where the characteristic time of collisional damping of Langmuir waves is much larger than the lifetime of type V emission.  Type V  polarizations are low (usually $< 0.07~\%$) which suggests harmonic emission.  However, it is common to find their polarization opposite in the sense of the corresponding type III.  \citet{Dulk_etal1980} suggest the most likely reason for this change is due to X-mode rather than O-mode emission.  This could be caused by increased isotropy in the Langmuir wave distribution as the condition for O-mode emission is Langmuir waves within 20$^o$ of the magnetic field.  Another deviation of type V emission from their associated type III emission is the occurrence of large position differences, sometimes up to 1 $R_s$ \citep{WeissStewart1965,Robinson1977}.  This is not always observed and the positions of the type III and V can overlap or only be slightly displaced. One explanation of type V emission is low energy electrons travelling along different magnetic field lines or a variation of the beaming of emission changes the position of the centroids \citep{Dulk_etal1980}.  Another explanation given for type V emission is the electron-cyclotron maser instability \citep{WingleeDulk1986,Tang_etal2013}.

\subsubsection{Type III-l bursts}

Associated with coronal mass ejections (CMEs) and solar energetic protons are type III-l bursts \citep{Cane_etal2002,MAcDowall_etal2003}.  These bursts start at much lower frequencies than normal  (hence the name typeIII-l), around 10 MHz and progress to lower frequencies as a function of time.  Moreover the bursts usually start delayed with respect to the associated flare and last on average 20 minutes.  It has been suggested \citep{Cane_etal2002} that type III-l bursts are produced by electrons accelerated by reconnection behind fast CMEs.  Conflicting accounts were found by \citet{Macdowall_etal2009,CliverLing2009} as the former found type III-l events were associated with intense SEP events favouring flare acceleration of the CME, whilst the latter found type III-l events associated with type II bursts favouring shock acceleration of the CME.  However, \citet{GopalswamyMakela2010} found that the presence of a type III-l burst does not always signify the presence of solar energetic protons.

\subsubsection{Type III storms}

A phenomena called type III storms can occur when type III bursts are observed quasi-continuously over a period of days \citep{FainbergStone1970a,FainbergStone1970b,FainbergStone1971}.  These type III bursts are narrow-band and usually occur at low frequencies $< 100$~MHz.  Type III storms are usually observed as harmonic emission with a low degree of polarisation.  However, fundamental-harmonic pairs are observed in type III storms with the fundamental component having a higher polarisation and is usually fragmented like a type IIIb burst \citep{DelanoeBoischot1972}.  Type III storms have been associated with type I noise storms, a similar phenomena at higher frequencies but with no frequency drift rate \citep[a good review of type I storms can be found by][]{Kai_etal1985} which in turn are associated with a solar active region \citep[e.g.][]{Kayser_etal1987,Gopalswamy2004}.  The polarisation degree of type III storms can remain constant during the entire length of the storm.  \citet{Reiner_etal2007} analysed storms below 1~MHz and found only a $5\%$ degree of polarisation, different from the higher $25\%$ polarisation degree found for type III storms above 20~MHz by \citet{Kai_etal1985}.  However, \citet{Reiner_etal2007} found the polarisation peaks near central meridian crossing of the corresponding radio sources and decreases systematically with decreasing frequency.  They deduce from these polarisation measurements that the coronal magnetic field (at least above the active regions they study) does not fall off as $1/R^2$ but slower.

\begin{SCfigure}
  \centering
  \caption{An example of a type III storm over a 6 hour period observed by the WAVES experiment onboard the WIND spacecraft \citep{Bougeret_etal1995}.  This type III storm went on for a number of days.}
  \includegraphics[width=0.49\columnwidth]{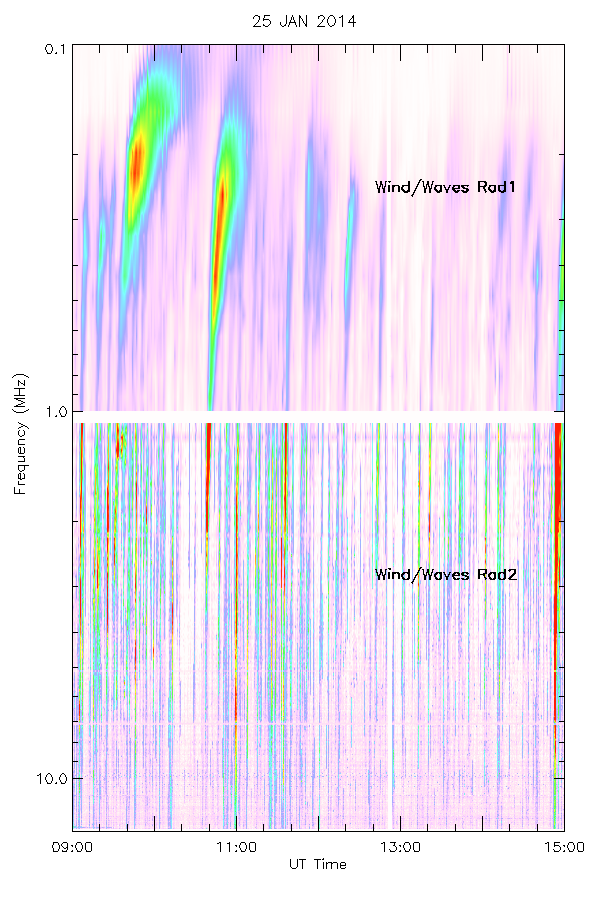}
\label{fig:typeIII_storm}
\end{SCfigure}

Type III storms usually consist of many faint type III bursts. Analysing the flux distribution at two specific frequencies 1.34~MHz and 860 kHz, \citet{Morioka_etal2007} found the occurrence rate of type III storm emission obeyed a power-law distribution with spectral index of -3.7 and -3.6 respectively.   Different spectral indices between -2.36 to -2.10 were found by \citet{Eastwood_etal2010} at a number of different frequencies between 3 and 10 MHz.  The discrepancy between the two ranges of indices is suggested by \citet{Eastwood_etal2010} to be due to the averaging of a few storms and the use of data binning in \citet{Morioka_etal2007}.   In comparison the spectral index of flux occurrence rate of type I emission has been found to be -3 \citep{MercierTrottet1997}.

\subsection{Multi-wavelength studies} \label{sec:HXRs}

\subsubsection{Type IIIs and X-rays} 

Type III radio bursts are not the only wavelength of light that can help diagnose the properties of accelerated electrons at the Sun.  X-ray emission is frequently observed during solar flares and is believed to also be driven by accelerated electrons.   Electrons travel into the dense solar chromosphere and thermalise via electron-ion Coulomb collisions.  Bremsstrahlung X-rays are emitted  \citep[see][as a recent review]{Holman_etal2011} and are detected both directly and through X-rays reflected from the solar surface \citep[e.g.][]{KontarJeffrey2010}.  We can use the X-ray signature to deduce the temporal, spatial, and energetic profile of the energised electrons \citep[e.g.][]{Kontar_etal2011}.

\begin{SCfigure}
  \centering
  \includegraphics[width=0.59\columnwidth]{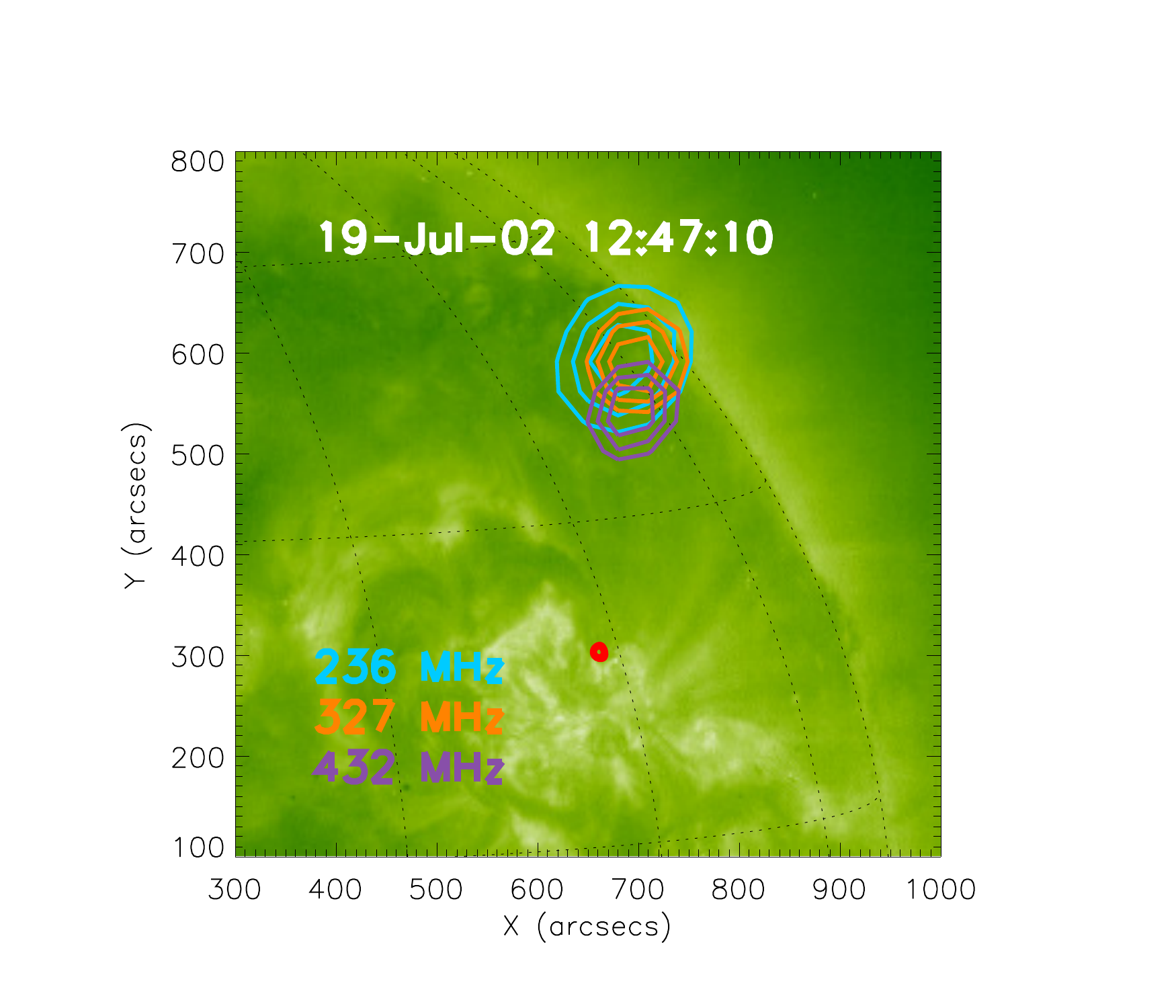}
  \caption{An example flare imaged by RHESSI \citep{Lin_etal2002} at 12-25 keV in X-rays (red) and by the Nan\c{c}ay Radioheliograph \citep{KerdraonDelouis1997} at 432 MHz (purple), 327 MHz (orange), and 237 MHz (blue).  All contours are at 10 \% intervals starting from 70 percent.  The background is an EIT $195~\text{\AA}$}
\label{fig:nrh_image}
\end{SCfigure}

The simultaneous observation of hard X-rays (HXR) and metric/decimetric radio emission is commonplace during flares and the relationship between type III bursts and hard X-ray emissions has been studied for many years \citep[see for example][for a review]{PickVilmer2008}.  An example flare showing simultaneous X-ray and type III emission is shown in Figure \ref{fig:nrh_image}.  A recent study by \citet{Benz_etal2005,Benz_etal2007} on 201 flares above GOES class C5 found that nearly all flares were associated with coherent radio emission between the range 4~Ghz to 100~MHz.  The majority of flares in the study that had coherent radio emission had a type III burst or groups of type III bursts.

There have been many statistical studies between coherent type III radio emission and HXR bursts \citep[e.g.][]{Kane1972,Kane1981,Hamilton_etal1990,Aschwanden_etala1995,ArznerBenz2005}.  Bursts of type III radio emission have been found to temporally correlate with bursts in HXRs \citep[e.g.][]{Aschwanden_etalb1995,ArznerBenz2005}.  However, there is often a small delay in the emission \citep[e.g.][]{Dennis_etal1984,Aschwanden_etala1995} of 2 seconds or less.   A type III/X-ray correlation is systematically more likely when the intensity of the HXR or radio emission increases \citep{Kane1981,Hamilton_etal1990}.  Such a property could be the reason for the high association rate of $>C5$ GOES class flares found by \citet{Benz_etal2005}.  A higher peak starting frequency of the type III bursts has also been found more likely in type III/HXR correlated events along with a low spectral index of HXRs (hard spectrum).

Recently, \citet{Reid_etal2014} found a correlation between the starting frequency of type III bursts and the hard X-ray spectral index in flares.  The hard X-ray spectral index is proportional to the accelerated electron beam spectral index.  A higher spectral index means the electron beam will become unstable faster and is thus able to generate type III radio emission at lower heights and consequently higher starting frequencies.  \citet{Reid_etal2011,Reid_etal2014} used the connection between type III starting frequency and HXR spectral index, first used by \citet{Kane_etal1982,Benz_etal1983}, to deduce the spatial characteristics of flare acceleration regions.  They found acceleration region starting heights varied between 25 to 200 Mm with a mean of 100 Mm.  The acceleration region vertical extent varied between 16 to 2 Mm with a mean of 8 Mm.  Some heights are in good agreement with the height found from time-of-flight X-ray analysis \citep{Aschwanden_etal1998} but others are higher in the corona.

It should also be theoretically possible to observe hard X-rays associated with the type III producing electron beams in the upper corona.  The background density is much lower and so emission is expected to be extremely faint \citep{Saint-Hilaire_etal2009}.  That we have not detected such X-ray emission yet, \citet{Saint-Hilaire_etal2009} use to estimate that the number of electrons escaping in type III producing electron beams must be $\leq 10^{34}$ electrons s$^{-1}$ above 10 keV.

The effect of producing Langmuir waves alters the distribution of electrons (see Section \ref{sec:theory}).  Such an effect has been studied in the context of flare accelerated electron beam propagating down through the corona and into the dense chromosphere to generate X-rays \citep{EmslieSmith1984,HamiltonPetrosian1987,McClements1987b,McClements1987a,Hannah_etal2009,ZharkovaSiversky2011,HannahKontar2011,Hannah_etal2013}.  If an electron beam is accelerated high enough in the corona it can become unstable to Langmuir wave growth and the spectral index of the electron beam can be modified.  The electron spectrum tends to flatten (decrease in magnitude) and the total amount of X-rays emitted is less.  For strong plasma inhomogeneity electrons can be re-accelerated to high energies (see Section \ref{sec:theory}) and the total amount of X-ray radiation emitted can be more than predicted with a simple collisional model \citep{Hannah_etal2013}.

\subsubsection{Type IIIs and EUV/X-ray jets} \label{sec:EUV}

The joint observations of type III bursts and soft X-ray (SXR) jets have been studied by a number of different authors \citep{Aurass_etal1994,Kundu_etal1995,Raulin_etal1996}.  They found the emission was co-spatial and the centroids of the type III bursts at different frequencies were aligned with the soft X-ray jet indicating that particle acceleration is taking place when coronal jets occur.  Jets that are correlated with type III bursts are also observed in EUV \citep[e.g.][]{Innes_etal2011,Klassen_etal2011,Chen_etal2013a,Chen_etal2013b} and are commonly associated with $^3$He-rich solar energetic electron events \citep{Wang_etal2006,Pick_etal2006,Nitta_etal2008}.  The jets in these studies frequently arose from the border of an active region and a coronal hole.  A study involving EUV and X-ray jets where simultaneous type III bursts were observed \citep{Krucker_etal2011} found hard X-ray footpoints that strongly suggest reconnection with a closed and open magnetic field.  Hard X-rays have recently been observed to accompany EUV jets \citep{BainFletcher2009,Glesener_etal2012} where type III bursts were also present.  This further strengthens the case that acceleration of particles is present at the same time that thermal plasma is ejected from the solar corona in the form of a jet.  Other evidence is the in-situ detection of electron spikes $<300$~keV \citep{Klassen_etal2011,Klassen_etal2012} at 1~AU which are very closely temporarily correlated with type III bursts and EUV jets although there is a slight timing issue (see Section \ref{sec:electrons}). 

A recent paper by \citep{Chen_etal2013b} has used the new upgraded EVLA \citep{Perley_etal2011} to make images of a type III burst at frequencies between 1 GHz and 2 GHz that occurred at the same time as an EUV jet and HXR footpoint observations.  From the images they infer the acceleration site of the type III bursts is below 15 Mm in the solar corona and the electron beam travels along bundles of discrete magnetic loops into the high corona.  Moreover, by using EUV images they deduce that the coronal loops the type III producing electron beams are travelling along may be cooler and denser than the surrounding plasma.  This dramatically helps the radio emission to escape and is similar to the scenario proposed by \citet{Benz_etal1992} for allowing decimetric emission to escape the corona.

\section{In-situ plasma observations} \label{sec:electrons}

\subsection{Electron beams}

The first in-situ observations of energetic particles \citep{vanAllen1965} opened up the non-electromagnetic window of flare accelerated particle observations.  Since then solar energetic electron events have been found to be closely related observationally \citep[e.g.][]{Ergun_etal1998,Gosling_etal2003,Krucker_etal2007} and theoretically to solar type III radio bursts, having a $99\%$ association rate \citep{Wang_etal2012}.

Impulsive electron events often extend to ~1 keV \citep{Lin_etal1996} with some even extending down to the ~0.1 to ~1 keV energy range \citep{Gosling_etal2003}.  The presence of such low energy electrons favours injection sites high in the corona at altitudes $\geq 0.1 R_{\odot}$ as Coulomb collisions in the low solar atmosphere will cause low energy electrons to lose their energy.  Electron time profiles generally show a rapid onset and also near time-of-flight velocity dispersion \citep[e.g.][]{Lin1985,Krucker_etal1999,Krucker_etal2007}.  Electrons also have a beamed pitch-angle distribution at lower energies $<20$~keV \citep[e.g.][]{Lin1990,Wang_etal2011} but can have very high pitch angle distributions at high energies of 300 keV.

A statistical study of 10 years of solar energetic electron events using the 3D plasma and Energetic Particle (3DP) instrument \citep{Lin_etal1995} on the WIND spacecraft was carried out \citep{Wang_etal2012} finding near 1200 events between 1995 and 2005.  They found an occurrence rate of solar electron events can be fitted as a power-law at 40 keV and 2.8 keV with a spectral index between 1.08 - 1.63 and 1.02 - 1.38 respectively, that changes every year.  This power-law is smaller than the one associated with type III radio bursts (1.7) observed at the high frequencies by \citet{Saint-Hilaire_etal2013}.  One possible reason could be due to the high electron background flux at low energies \citep{Wang_etal2012} obscuring events that could steepen the power-law dependence

Solar impulsive electron events detected in-situ generally display broken power-law energy distributions with lower energies having harder spectra.  Broken power-law distributions have been observed for some time \citep[e.g.][]{Wang_etal1971} but their origin has remained ambiguous, being either a signature of the acceleration mechanism or a transport effect.  A recent statistical survey was carried out by \citet{Krucker_etal2009} on 62 impulsive events.    They found the average break energy was $\approx 60$~keV with averaged power-law indices below and above the break of $\delta_{low} = 1.9 \pm 0.3$ and $\delta_{high} = 3.6 \pm 0.7$ respectively.  The power-law indices have an average ratio $\delta_{low}/\delta_{high}$ of 0.54 with a standard deviation of 0.09.  The power-law indices also correlate with a coefficient of 0.74.  The presence of the broken power-law can be replicated by simulating electron beam propagation to 1~AU and their interactions with Langmuir waves (see Section \ref{sec:theory}).

It is often believed that solar energetic electron events propagate scatter-free from the Sun to the Earth \citep[e.g.][]{Wang_etal2006}.  The observed correlation between the spectral indices of energetic electrons at the Sun from X-ray data and the Earth from in-situ data  \citep{Lin1985,Krucker_etal2007} is often viewed as an additional support for scatter-free transport.  However, the injection time of energetic electrons at different energies does not correspond to the inferred injection time by looking at the associated type III bursts \citep{Krucker_etal1999,HaggertyRoelof2002,Cane2003,Wang_etal2006,Wang_etal2011,Kahler_etal2011b,Klassen_etal2012}.  Low energy electrons (e.g. 10 keV) have an early arrival time while high energy electrons (e.g. 100 keV) have a late arrival time.  The early arrival time could be due to pitch angle scattering \citep[e.g.][]{Tan_etal2011} or electron interaction with Langmuir waves \citep{KontarReid2009}.  It is not entirely clear what causes the later arrival time of the high energy electrons although an increased path length greater than 1~AU has been suggested \citep[e.g.][]{Kahler_etal2011b}.

\subsection{Langmuir waves}

In-situ observations of Langmuir waves associated with type III radio bursts were first taken by \citet{GurnettAnderson1976,GurnettAnderson1977} using the Helios spacecraft at around 0.5~AU.  They found that the distribution of Langmuir waves is very clumpy in space.  Observations at 1~AU came later using the ISEE-3 spacecraft \citep{Lin_etal1981}.  More recent observations of clumpy Langmuir waves show similar properties \citep[e.g.][]{Kellogg_etal2009} using the time domain sampler in the WAVES instrument onboard the STEREO spacecraft \citep{Bougeret_etal2008}.  Recently \citet{Ergun_etal2008} modelled the localisation of Langmuir wave packets as a superposition of Langmuir eigenmodes and this concept has been worked on by other authors \citep[e.g.][]{GrahamCairns2013b}.   There has been a number of recent works claiming to observe Langmuir wave collapse in the solar wind \citep[e.g.][]{Thejappa_etal2012,Thejappa_etal2013}.  However, other studies have analysed the same events but using all three components of the electric field \citep{Graham_etal2012a,Graham_etal2012b} and find that the electric field is too weak for wave collapse to occur.  Recently \citet{Vidojevic_etal2011,Vidojevic_etal2012} analysed the statistical distribution of Langmuir waves measured near 1~AU by the Wind spacecraft \citep{Bougeret_etal1995}.  By looking at the Langmuir wave distributions of 36 different electron beams with careful background subtraction they found that the distribution of Langmuir waves are more accurately modelled by a Pearson's system of distributions \citep{Pearson1895} and not a log-normal distribution as previously thought.

\subsection{Solar Wind Conditions}

The plasma of the solar corona and the solar wind is a non-uniform, turbulent medium with density perturbations at various length scales \citep[see e.g. ][as a review]{BrunoCarbone2005}.  Observations of how this background plasma fluctuates is important for modelling the Langmuir wave generation of electron beams responsible for type III radio bursts.  

In-situ measurements have been used to determine the density spectrum near the Earth and between $0.3$ and $1$~AU with {\it Helios} \citep{MarschTu1990}. The spectral slope at frequencies below $10^{-3}$~Hz were found to have a tendency to get smaller the closer the spacecraft got to the Sun in the fast solar wind.  These results were further extended by \citet{Woo_etal1995} using Ulysses remote sensing radio measurements for distances $<40~R_s$ which predicted the decrease in r.m.s. deviation of the density turbulence in the fast solar wind at wavenumber $k=1.4\times10^6~\rm{km}^{-1}$.  The results for the slow solar wind density turbulence \citep{MarschTu1990,Woo_etal1995} showed a constant level around $10~\%$ which was also found in the later study from \citet{Spangler2002}.  More recently \citet{Chen_etal2012,Chen_etal2013c} have examined the slope of the density fluctuations power spectral density for 17 intervals of solar wind.  They find a -5/3 power-law below $0.1~k\rho_i$ where $\rho_i$ is the proton gyroradius.  Between 0.1 and 1 $k\rho_i$ the spectrum flattens to -1.1 in-line with other observations at 1~AU \citep[e.g.][]{Celnikier_etal1983,KelloggHorbury2005} and closer to the Sun \citep{ColesHarmon1989,Coles_etal1991}.  Between 3 and 15 $k\rho_i$ \citet{Chen_etal2013c} report a steepening of the spectral index to -2.75.

Background solar wind electrons are not in thermal equilibrium with a Maxwellian distribution but exist in a quasi-thermal state, with electrons extending to much higher energies \citep{Lin_etal1972}.  Their velocity distribution function at all pitch angles is usually modelled using two convecting bi-Maxwellians, the core and the halo.  A skewed distribution also exists in the fast solar wind parallel to the magnetic field direction.  Known as the strahl, this high energy tail usually propagates away from the Sun and has a narrow pitch angle distribution between 10-20$^o$ wide.  Observations of the background solar wind electrons have shown that a kappa distribution can better model the solar wind \citep{Maksimovic_etal2005,Lechat_etal2010}.  The kappa distribution more accurately models the electron temperature whilst having fewer free parameters than the sum of two Maxwellians.  Recently \citep{LiCairns2014} have used a kappa distribution when modelling type III radio bursts.  They found that a kappa distribution leads to a type III burst that drifts with a faster velocity.  The increased level of higher energy background electrons in the kappa distribution reduce the Langmuir waves that are generated by slower velocity electrons.  As such we only observe radio emission generated by higher energy electrons (and hence higher velocity) in the beam.  The corresponding type III burst has a higher drift rate in frequency space.

\section{Electron beam interaction with Langmuir waves} \label{sec:theory}

The current generally accepted mechanism for type III burst generation is the plasma emission mechanism as described in the introduction. Since the original proposal of \citet{GinzburgZhelezniakov1958}, large amounts of work have been invested in the development of this theory.  However, due to its complicated nature the exact details are still not fully understood. Many analytical treatments of the various steps exist and these have been supplemented by numerical simulations in order to develop the general picture of type III burst production as presented here. 

The generation process is summarised in Figure \ref{fig:typeIII_flow}. In this section we discuss the electron propagation and the resulting Langmuir wave generation.  The evolution of the Langmuir waves via non-linear scattering and wave-wave interactions is the topic of Section \ref{sec:em} along with the processes that lead to second harmonic emission (at twice the local plasma frequency) and fundamental emission (first harmonic) at the local plasma frequency.

\subsection{Quasi-linear theory}

The rapid frequency drift (see Section \ref{sec:freqtime}) of type III bursts is caused by the rapid motion of the generating electrons, and can be derived from a prescribed plasma density profile by assuming that the electrons stream freely \citep[e.g.][]{Karlicky_etal1996, RobinsonBenz2000,Ledenev_etal2004}. Alternatively, assuming a constant velocity for the exciting electrons we can derive the density profile of the ambient plasma \citep[e.g.][]{2009ApJ...706L.265C}.  However, the electrons do not stream freely but produce Langmuir waves. Early numerical simulations by e.g. \citet{TakakuraShibahashi1976,Takakura1982}, showed that Langmuir wave generation must be considered to reproduce the electron behaviour over long distances, and therefore even the basic properties of type III bursts.

The quasi-linear equations describing the interaction of electrons with Langmuir waves were introduced by \citet{Vedenov_etal1962,DrummondPines1962}. The 1-D version of these is often employed in relation to type III bursts, as we may assume the generating electrons are closely tied to the solar magnetic field, as has been observed in-situ near the Earth (Section \ref{sec:electrons}). Additional effects are seen in 2-d modelling \citep[e.g.][]{Ziebell_etal2008a,Ziebell_etal2008b,Pavan_etal2009a,Pavan_etal2009b,Ziebell_etal2011,Ziebell_etal2012} such as heating of the background plasma \citep[e.g.][]{Yoon_etal2012,Rha_etal2013}. However the 1-d dynamics are generally expected to dominate.

Writing the electron distribution function as $f(v,t)$ [electrons cm$^{-4}$ s], normalised so that $\int f(v,t) dv =n_e$ with $n_e$ the plasma density in cm$^{-3}$ and the spectral energy density of Langmuir waves as $W(v,t)$ [ergs cm$^{-2}$] with the normalisation $\int W(v,t)\, dv=E_L$ the total energy density of the waves in erg cm$^{-3}$ we have

\begin{equation}
\frac{\partial f(v,t)}{\partial t} =
\frac{4\pi^2e^2}{m_e^2}\frac{\partial }{\partial v}\frac{W}{v}\frac{\partial f(v,t)}{\partial v}
\label{eqk1}
\end{equation}
\begin{equation}
\frac{\partial W}{\partial t} = \frac{\pi \omega_{pe}}{n_e}v^2W\frac{\partial f(v,t)}{\partial v}
\label{eqk2}
\end{equation}
where $v$ signifies the electron kinetic velocity and the Langmuir wave phase velocity, as these must be equal for interaction to occur. This is equivalent to the Cherenkov resonance condition that the wave wavenumber and frequency $k, \omega$ and the particle velocity $v$ must satisfy $\omega = k v$.

The quasilinear equations are derived from the Vlasov equation by ignoring all other electromagnetic processes present in the plasma such as collisions.  Several assumptions are needed, such as the absence of particle trapping, and the non-magnetisation of the electrons, so that their gyroradius is large and their trajectories almost linear. We also need the weak-turbulence condition which states that the energy in Langmuir waves is much less than that in the background plasma. Provided the perturbations created on a particle through wave-particle interactions are small (for example much less than an electron gyroradius in a gyroperiod) the quasilinear equations are valid.

The growth rate of waves is (amongst other things) proportional to $\partial f/\partial v$ (Equation (\ref{eqk2})).  When the electron distribution has a positive gradient in velocity space we excite waves and vice versa.  An intuitive way to this about this is as follows.  The energy exchange between the particles and waves occurs for particles with speeds $v$ close to $\omega/k$; particles with $v<\omega/k$ gain energy from the waves while particles with $v>\omega/k$ lose energy from the waves.  The waves gain energy if we have more particles satisfying $v>\omega/k$ and the waves lose energy if we have more particles satisfying $v<\omega/k$.  

The effect of wave generation on the electron distribution is more than simple energy loss. The right hand side of Equation (\ref{eqk1}) is a diffusion equation,  with the diffusion operator $D=W/v$. The transfer of energy from electrons to waves and back to the electrons causes the electrons to diffuse in velocity, smoothing out the positive gradient and slowing the energy transfer. The asymptotic (long time) solution to the quasi-linear equations is thus a plateau in velocity space \citep[e.g.][]{VedenovRyutov1972,Grognard1985,Kontar2001a}.

\begin{figure}[tp]
  \centering
  \includegraphics[width=0.79\textwidth]{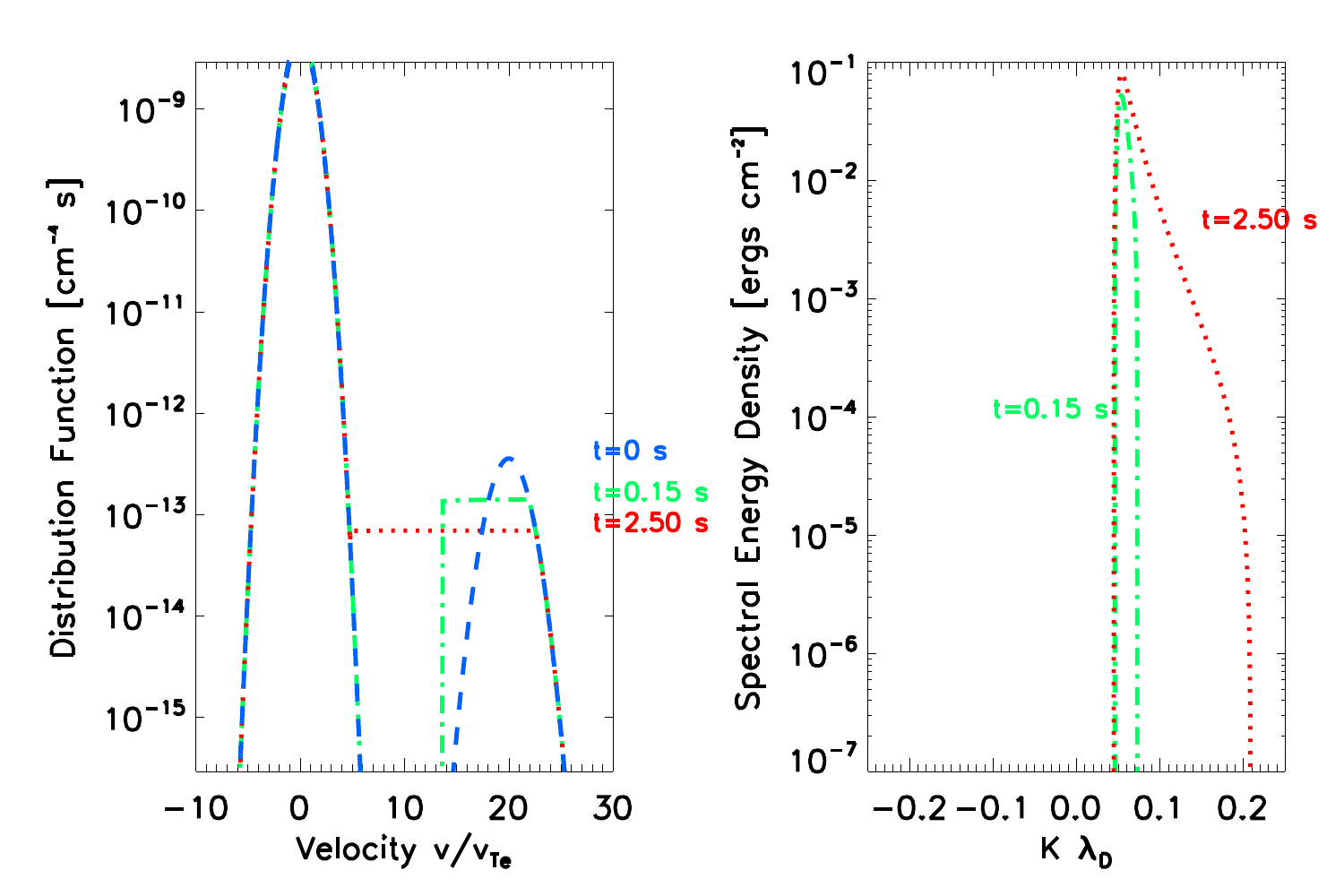}
  \caption{A demonstration of an electron beam inducing Langmuir waves in the background plasma.  Left: the electron distribution function of a background Maxwellian and an electron beam in velocity space (normalised by the thermal velocity $v_{Te}$.  The density ratio is  $n_b/n_e=10^{-4}$, $v_{Te}=5.5\times10^{8}~\rm{cm~s}^{-1}$.  Right:  the spectral energy density of Langmuir waves where the x-axis is normalised by the Debye length.  The three different colours correspond to different points in time.}
\label{fig:WPI}
\vspace{20pt}
\end{figure}

Figure \ref{fig:WPI} shows an example of the resonant interaction of Langmuir waves with electrons, and the resulting quasi-linear relaxation. The left panel shows the electron distribution consisting of a background Maxwellian plasma (density $n_e$) and an electron beam (density $n_b$) with a density ratio of $n_e/n_b=10^{-4}$.  The right panel shows the spectral energy density of Langmuir waves. This initial beam has positive gradient in velocity, and so is unstable to Langmuir wave growth. After a short time (0.15~s) the electrons have excited a high level of Langmuir waves, and the beam has diffused in velocity, becoming significantly wider in velocity space. At a later time (2.5~s) the electron distribution has relaxed to a plateau with zero gradient in velocity space, and the instability has therefore saturated. The level of induced Langmuir waves has grown, as has their width in wavenumber space. 

\subsection{Electron Beam Instability Distance}

The electrons accelerated during a solar flare are generally expected to have a power-law like distribution in velocity with a negative index.  Such a distribution will not generate Langmuir waves.  However, time-of-flight effects, where fast electrons outpace slower ones, produce a distribution with positive slope which is unstable to Langmuir wave generation.  This occurs only after the electrons have travelled a sufficient distance.  This instability distance is closely related to the starting frequency of type III bursts.  When the instability distance is small the electron beam will produce Langmuir waves (and hence radio emission) close to the dense acceleration site resulting in a high burst starting frequency.  When the instability distance is large, the electron beam travels into much lower density plasma before exciting Langmuir waves and so produces type III emission that starts at lower frequencies.  


\citet{Reid_etal2011} modelled the distance that an instantaneously injected electron beam would travel for a range of parameters.  They found that the size of the acceleration region and the spectral index of the electron beam were the dominating factors in calculating the starting frequency of the type III burst given a fixed height for the acceleration region.  Low spectral indices and small acceleration regions reduce the electron beam instability distance.  Such a dependency was used by \citet{Reid_etal2011,Reid_etal2014} to deduce heights and sizes of flare acceleration regions (see Section \ref{sec:HXRs}).  If the electrons are injected over a finite time (rather than instantaneously), the starting frequency will be affected \citep{ReidKontar2013}. The longer the injection time, the further the beam must travel to become unstable.

\subsection{Electron Beam Persistence}

In the early 60s, \citet{Sturrock1964} noted that without something stopping the beam-plasma instability, the rate of Langmuir wave generation by an electron beam in the solar corona was enough that the beam would lose all energy after propagating only metres. However type III burst observations showed that beams could persist to distances of 1~AU. The \emph{beam-plasma structure} was proposed to overcome this: here the electrons propagate accompanied by Langmuir waves which are continually generated at the front of the beam and reabsorbed at the back \citep[e.g.][]{ZheleznyakovZaitsev1970, Zaitsev_etal1972,Melnik1995}. Numerical work by e.g. \citet{TakakuraShibahashi1976,MagelssenSmith1977, Kontar_etal1998,Melnik_etal1999} confirmed that the beam-plasma structure could resolve the problem. 
	
Other processes can suppress the beam-plasma instability and therefore the Langmuir wave growth. In particular, density fluctuations in the plasma can shift Langmuir waves out of resonance with the electrons \citep[e.g.][]{SmithSime1979,Muschietti_etal1985}. Because the growth rate of Langmuir waves is proportional to their current level at a given wavenumber, this can suppress the instability \citep[in line with][]{Kontar2001d,Ledenev_etal2004,Li_etal2006b,ReidKontar2010}. The density fluctuations can be either wave modes with wavelengths comparable to the Debye length \citep{Vedenov_etal1967,GoldmanDubois1982,Yoon_etal2005}, or longer wavelength inhomogeneities \citep{Ryutov1969,NishikawaRyutov1976,SmithSime1979,Kontar2001c,Kontar_etal2012}. In the latter case, the Langmuir wave evolution can be described in the WKB approximation by \citep[e.g.][]{Ryutov1969,Kontar2001d}
\begin{equation}
\frac{\partial W}{\partial t} - \frac{\partial \omega_{pe}}{\partial x}\frac{\partial W}{\partial k} = 0.
\end{equation}
and as the plasma frequency is $\omega_{pe} \propto \sqrt{n_e}$, even weak density gradients are seen to strongly affect Langmuir waves. 

For a propagating beam-plasma structure, density fluctuations will lead to energy loss, because when Langmuir waves are shifted out of resonance with the beam their energy can no longer be reabsorbed. Recent numerical simulations \citep{KontarReid2009,ReidKontar2010,ReidKontar2013} taking into account (amongst other processes) the refraction of Langmuir waves, found that the beam does lose energy but is still able to persist over distances of 1~AU or more. Additionally, a power-law electron spectrum injected at the Sun was found by \citet{KontarReid2009} to be modified by density fluctuations and further from the Sun is better modelled as a broken power law \citep{ReidKontar2013}. This is due to the combination of Langmuir wave generation and their spectral evolution. Additionally, the spectral evolution of Langmuir waves to small wavenumbers can in some circumstances lead to acceleration of high-velocity electrons in the tail of the electron distribution, because the Langmuir waves can transfer energy from the beam electrons into this tail. This effect has been studied numerically by  e.g. \citet{Kontar2001d,ReidKontar2010,Ratcliffe_etal2012,VoshchepynetsKrasnoselskikh2013,Krafft_etal2013}.

\subsection{Langmuir Wave Clumping} \label{sec:el_sims}

Plasma density fluctuations are also believed to be responsible for the clumpy Langmuir wave distribution observed in-situ near the Earth.  Background density fluctuations are known to correlate to the clumpy observations of Langmuir waves in the solar wind \citep{Robinson_etal1992}.  Recent simulations of the effects of density fluctuations on type III emission were carried out by \citet{Li_etal2012}. They considered a plasma density profile with several density enhancements and found these produced strong fine structure in both the fundamental and the harmonic component that resemble stria bursts.  However, the fundamental component in these simulations is very weak and thus unobservable.  The effect of temperature variation has also been investigated both for the electron temperature \citep{2011ApJ...730...20L} and the ion temperature \citep{2011ApJ...730...21L}.  Variations only in electron temperature \citep{2011ApJ...730...20L} produced similar results to those presented for density fluctuations although in general the emission was less intense.  Interestingly, large fluctuations in the ion temperature \citep{2011ApJ...730...21L} were able to produce more distinct fluctuations in the fundamental component and flux levels were able to reach 1~sfu.  It is desirable to observe fragmented fundamental emission because the polarisation of type IIIb bursts is usually observed to be quite high (see Section \ref{sec:polarisation}).  The characteristics of the fine structures in these simulations resemble striae and thus density and/or temperature fluctuations are a promising explanation, although they cannot as yet explain the occurrence of normal H and modulated F pairs.

An alternative way to deal with density fluctuations was introduced by \citet{Robinson1992,Robinson_etal1992} called `Stochastic Growth Theory' (SGT).  The main principle behind SGT is that density fluctuations induce random growth of Langmuir waves. The bursts of Langmuir waves have an amplification factor of $\exp G$ where $G$ is a random function with zero mean value and non-zero standard deviation.  This predicts a log-normal statistical distribution of the electric field amplitude.   Stochastic Growth Theory has been used with a variety of different simulation techniques \citep[e.g.][]{RobinsonCairns1998,CairnsRobinson1999,Li_etal2006b,Li_etal2006c} and has produced simulation results that resemble the clumpy observations of Langmuir waves and synthetic spectra of type III bursts.  

Whilst quasilinear theory is very good at capturing the large scale effects of the electron beam - Langmuir wave interaction, it struggles to capture the small scale effects of single wave packets.  An alternative approach can be used to model these small scales using the Zakharov equations \citep{Zakharov1972}.  This approach models the electric field with an additional source term to describe the beam particles \citep[e.g.][]{Zaslavsky_etal2010,Krafft_etal2013}.  Modelling in this way is able to replicate wave clumping and thus strongly varying electric fields that has been observed in the solar wind \citep[e.g.][]{Ergun_etal2008}.  It is also able to describe the reflection of Langmuir waves in density cavities where small scale lengths are important and the quasilinear approach is invalid.

\section{Non-linear processes and Langmuir wave evolution} \label{sec:em}

As discussed in the introduction, many years of work have been invested in understanding the plasma emission mechanism. However, due to its complex nature, and the many interconnected processes, there are still many unanswered questions. One key feature of the mechanism is the non-linear processes which control the evolution of the Langmuir waves, and their conversion into radio waves.  While their details are complicated, the basic equations provide insight into emission properties such as brightness temperature and angular distribution and show us what plasma properties are important (e.g. temperature ratio).  In this section we will present the basic theory of these processes. The derivation of the relevant equations is complicated, and is given in detail in the books by \citet{1980MelroseBothVols,1995lnlp.book.....T} and in many papers. We give here only a short introduction, and then focus on the application of the non-linear equations to the problem of type III burst generation. 

\subsection{The general form of the non-linear equations}

The density fluctuations in the previous section were generally considered to have scale lengths larger than the Langmuir wavelength. However, density-perturbing wave modes such as ion-sound waves can have wavelengths comparable to Langmuir waves. In this case, we can describe their effects as wave-wave interactions using a set of non-linear equations. The scattering of Langmuir waves by individual plasma particles is also described using non-linear equations. Both of these processes are vital for the further evolution of beam-generated Langmuir waves. 

The key principle is ``detailed balance'': in short we have interactions of the form $\sigma \rightleftarrows \sigma' +\sigma''$ where $\sigma, \sigma'$ and $\sigma''$ denote wavemodes, which can be the same, or different, and/or $\sigma +i \rightleftarrows \sigma' +i'$, where $i, i'$ are an initial and scattered plasma ion.  For every wave we lose on the left-hand side we must gain a wave on the right-hand side. Using this principle we can write down the rates of change of the ``occupation number'' of waves in mode $\sigma$. This is given by\footnote{This definition varies slightly between authors. In \citet{1980MelroseBothVols} a factor $\hbar$ appears explicitly, and the $(2\pi)^3$ is omitted with corresponding changes in the quoted probabilities. \citet{1995lnlp.book.....T} uses the given definition.} $N_\sigma=(2\pi)^3 W_\sigma /\omega_\sigma$ where $W_\sigma$ is the spectral energy density.  The probability of an interaction is then calculated from the detailed plasma physics, and requires that we specify the exact modes involved. 

Writing $\vec{k}, \omega, \vec{k'},\omega'$ for the wavevector and frequency of mode $\sigma, \sigma'$ respectively, the ion scattering process $\sigma +i \rightleftarrows \sigma' +i'$ is described by the equation 
\begin{equation}\label{eq:IS3D} \frac{ d W_\sigma(\vec{k})}{d t}=\int \frac{d \vec{k}'}{(2\pi)^3} w_i^{\sigma\sigma'}(\vec{k}', \vec{k})\left[\frac{\omega}{\omega'}W_{\sigma'}(\vec{k}')-W_\sigma(\vec{k})-\frac{(2\pi)^3}{k_BT_i}\frac{\omega-\omega'}{\omega'}W_\sigma(\vec{k})W_{\sigma'}(\vec{k}')\right] \end{equation}
with probability $w_i^{\sigma\sigma'}$ depending on the wave modes involved, and a second equation with same probability and $\sigma$ and $\sigma'$ interchanged. The plasma ion absorbs the momentum change between the initial and final waves, and in the cases considered here this change is sufficiently small that we may neglect the effect on the ions. 

For the process, $\sigma \rightleftarrows \sigma' +\sigma''$ the equations take the form

\begin{equation}\label{eq:3Wv3D1}\frac{d W_\sigma(\vec{k})}{dt}=\omega \int\int d\vec{k}'d\vec{k}'' w_{\sigma\sigma'\sigma''}(\vec{k}, \vec{k}', \vec{k}'')\left[\frac{W_{\sigma'}(\vec{k}')}{\omega'}\frac{W_{\sigma''}(\vec{k}'')}{\omega''}-\frac{W_{\sigma}(\vec{k})}{\omega}\left(\frac{W_{\sigma'}(\vec{k}')}{\omega'}+\frac{W_{\sigma''}(\vec{k}'')}{\omega''}\right)\right]\end{equation}
where $w_{\sigma\sigma'\sigma''}(\vec{k}, \vec{k}', \vec{k}'')$ is the emission probability. Equations for the other wave modes are similar, having the same probability but some sign changes. 
 
 The growth of mode $\sigma$ occurs due to the product term (the first term in the square brackets)
and the decay of this mode depends on both its own energy density and those of the decay products, giving the other two terms.
So for example, if we start with a high level of mode $\sigma$ and thermal levels of $\sigma', \sigma''$, that is, the level generated spontaneously by plasma particles, the interaction $\sigma \rightleftarrows \sigma' +\sigma''$ will proceed to the right, generating the modes on the right-hand side, from that on the left hand side. Eventually, we can reach the state where $d W_\sigma /dt=0$ and the interaction is saturated. 

Along with Equation (\ref{eq:3Wv3D1}), we have some conservation conditions. Energy conservation in this equation may be expressed by the condition \begin{equation}\label{eq:consOm}\omega^\sigma(\vec{k})=\omega^{\sigma'}(\vec{k}')+\omega^{\sigma''}(\vec{k}'')\end{equation} where $\omega^\sigma$ is the frequency of a wave in mode $\sigma$, and is included to emphasize that we must consider the relevant dispersion relation for the mode in question. Momentum conservation is given by \begin{equation}\label{eq:consK}\vec{k}= \vec{k}'+ \vec{k}''\end{equation} for $\vec{k}, \vec{k}', \vec{k}''$ the respective wave vectors. For a 3-wave interaction to be possible we must be able to simultaneously solve energy and momentum conservation using the relevant dispersion relations (frequency-wavevector relations), e.g. those for Langmuir and EM waves: \begin{equation} \label{eq:dispL}\omega_{L}(\vec{k}) \simeq \omega_{pe} + 3v_{Te}^2k^2/(2\omega_{pe}) \,,\; \omega_{EM}(\vec{k})=(\omega_{pe} + c^2k^2)^{1/2}\end{equation} respectively. Thus for example, the interaction $L + s \rightleftarrows  L' $ is allowed, but $L + L' \rightleftarrows L''$ is not as the resulting wave vector must be larger than $k_{De}$, and this is not possible for Langmuir waves. 

\subsection{Non-linear Langmuir wave evolution}

For Langmuir wave evolution, the probability used in Equation (\ref{eq:IS3D}) for the scattering $L + i \rightleftarrows L' + i'$ is
\begin{equation}\label{eq:cIon} w_i^{L L'}= \frac{\sqrt{\pi} \omega_{pe}^2}{2 n_e v_{Ti} (1+T_e/T_i)^2} \frac{|{\vec{k}} \cdot {\vec{k}'}| }{k k' |\vec{k}-\vec{k}'|} \exp{\left(-\frac{(\omega'-\omega)^2}{2|\vec{k}'-\vec{k}|^2 v_{Ti}^2}\right)}. \end{equation} 
Due to the dot product, this probability is dipolar, i.e. it is maximised for $\vec{k}'$ either parallel or antiparallel to $\vec{k}$. The exponential term implies 
$\omega' \simeq \omega$ and so $|\vec{k}'|\simeq |\vec{k}|$. Finally,  the appearance of $|\vec{k}-\vec{k}'|$ in both the exponential and main term implies that we mainly get backscattering, so that $\vec{k}' \simeq - \vec{k}$. 

In Equation (\ref{eq:IS3D}), the first two terms in the square brackets may be called spontaneous scattering, and give approximately linear growth or decay. Conversely, the third term depends on the spectral energy density already present. Because of the factor $\omega-\omega'$, the scattered waves will grow where $\omega' < \omega$, and decay where $\omega' > \omega$. Thus the main growth occurs for $\vec{k}' \simeq - \vec{k}+ \Delta k$ with the small decrement $\Delta k$. 

The main three-wave process affecting the Langmuir waves is $L \rightleftarrows L' + s$ where $s$ denotes an ion-sound wave, with the approximate dispersion relation \begin{equation}\label{eq:disps}\omega = k v_s  \, ,\; v_s=\sqrt{\frac{k_B T_e}{M_i} \left(1+\frac{3 T_i}{T_e}\right)},\end{equation} the sound speed. The probability for this process which appears in Equation (\ref{eq:3Wv3D1}) is \begin{equation}w_{LLs}(\vec{k}, \vec{k}', \vec{k}'') =\frac{\pi \omega_{pe}^2}{4 n_e T_e}\left(1+\frac{3T_i}{T_e}\right)\omega_k^s \left(\frac{(\vec{k}'\cdot\vec{k}^{\prime\prime})^2}{k^{\prime 2}k^{\prime\prime
2}}\right)\delta(\omega_{k'}^l-\omega_{k''}^l-\omega_k^s)
\end{equation}

Solving the energy and momentum conservation equations (Equations (\ref{eq:consOm}) and (\ref{eq:consK}) simultaneously using the appropriate wave dispersion relations tells us that the product Langmuir wave has a wavevector approximately anti-parallel to the initial one i.e.  $\vec{k}'' \simeq -\vec{k}'$. The ion-sound wave must then have $\vec{k} \simeq 2\vec{k}'$. Thus as in the previous case of ion-scattering, we mainly generate backscattered Langmuir waves. 

Again, rather than exact backscattering, the final Langmuir wavenumber has a slight decrement, so we have ${k}''=-{k}' +\Delta k $ with \begin{equation}\label{eq:deltaK} \frac{\Delta k}{k_{De}}=\frac{2}{3}\sqrt{\frac{m_e}{m_i}}\sqrt{1+\frac{3T_i}{T_e}} \sim \frac{1}{30}.\end{equation} Thus repeated Langmuir wave scatterings produce waves at smaller and smaller wavenumbers, and can lead to significant spectral evolution. This can be important for the electromagnetic emission processes as there can be strong constraints on the participating wavenumbers. 
 
A thermal level of ion-sound waves is naturally present in plasma and the high levels of Langmuir waves generated by an electron beam can produce rapid growth of ion-sound waves and scattered Langmuir waves. However, ion-sound waves are subject to Landau damping with damping coefficient \begin{equation}\gamma_k^s=\sqrt{\frac{\pi}{8}}\omega_k^s\left[ \frac{v_s}{v_{Te}}+\left(\frac{v_s}{v_{Ti}}\right)^3\exp{-\left(\frac{v_s}{v_{Ti}}\right)^2}\right].
\end{equation} In plasma with $T_i \simeq T_e$ this is very strong, approaching the ion-sound wave frequency, and thus the three-wave processes become inefficient. Referring to Equation (\ref{eq:3Wv3D1}), we see that this is because the positive ``driving'' term for the decay process is proportional to $W_s$ and with strong damping this quantity remains small. 

\subsection{Harmonic Emission Equations}

Because the temperature ratio in the solar corona and wind can vary \citep[e.g.][]{1998JGR...103.9553N, 1979JGR....84.2029G} from $T_i/T_e \lesssim 0.1$ to $1$ or even 2, the ion-scattering and ion-sound wave decay processes described in the previous section vary in efficiency, with the latter dominating where the ratio is lower, and the former where it is large. However, in both regimes, we can produce a large level of Langmuir waves oppositely directed to the initial beam-generated population, and thus efficiently produce radiation at the second harmonic of the plasma frequency.

The growth rate for harmonic emission by the process $L+L' \rightleftarrows t$ is again given by Equation (\ref{eq:3Wv3D1}), with a probability \begin{equation}\label{eqn:LLTProb}
w^{LLT}(\vec{k}_1, \vec{k}_2,\vec{k}_T)=\pi \omega_{pe} \frac{(k_2^2-k_1^2)^2 (\vec{k}_T\times\vec{k}_1)^2}{16 m_e n_e k_T^{2}k_1^2k_2^2}\delta(\omega_{k_T}^{T}-\omega_{k_1}^L-\omega_{k_2}^L)
\end{equation} where we have labelled the participating Langmuir wavevectors as $k_1, k_2$ for clarity. 

Using the energy and momentum conservation conditions (Equations (\ref{eq:consOm}) and (\ref{eq:consK}) and the dispersion relations (Equation (\ref{eq:dispL})) we find that the coalescing Langmuir waves must be approximately oppositely directed. This is referred to as the ``head-on approximation'' (HOA), and allows us to simplify the probability, by replacing $(k_2^2-k_1^2)^2/k_2^2 = (\vec{k}_T \cdot \vec{k}_1)^2/ k_1^2$. 

To derive this, we must assume that $k_T \ll k_1, k_2$.  For typical beam electrons, with $v \simeq 10 v_{Te}$ the generated Langmuir waves have $k \simeq 0.1 k_{De}$ and from ${(\omega_{pe}^2 +c^2 k_T^2)^{1/2}}\simeq 2 \omega_{pe}$ we have that $k_T \simeq \sqrt{3} \omega_{pe} /c$. The ratio $k_T/k_1$ is then of the order 0.2, which is not very small. Clearly then the coalescence is not perfectly head on. 

Some simple calculations using Equation (\ref{eqn:LLTProb}) can show that even accounting for the angle between $k_1$ and $k_2$, the most probable angle between $k_1$ and $k_T$ remains close to 45$^o$. More detailed calculations by e.g. \citet{1979A&A....73..151M,1996PhPl....3..149W,1997SoPh..171..393W} used specific angular distributions for the Langmuir waves to find the emission probability with and without applying the head-on approximation. Significant differences were found at small wavenumbers. Thus the detailed study or simulation of harmonic emission is complicated. It must either be treated using a fully 3-dimensional model, or by analytical angle averaging. 

There are however two great simplifications which may be made to the equations for harmonic emission without affecting the results obtained. The energy lost from the Langmuir waves is very small, and so can be neglected when considering the Langmuir wave evolution. The high group velocity of the EM waves, given by \begin{equation}\label{eq:vg} \vec{v}_g = {c^2 \vec{k}}/{\omega} \end{equation} and approximately equal to $c$ for emission at $2 \omega_{pe}$ means the emission rapidly leaves the source region where Langmuir waves are present. Thus the reverse process of $t \rightarrow L + L'$ can often be ignored, leaving only a growth rate of $d W_{harmonic} /dt \propto W_L W_L'$ for $W_L, W_L'$ the beam generated and backscattered Langmuir wave populations. Balancing this emission rate with the propagation losses from the source we can find the approximate temperature where the emission saturates. This depends on the beam and plasma parameters as well as the source model, but is usually many orders of magnitude above thermal and easily able to explain the observed brightness temperatures.

\subsection{Fundamental Emission Equations}

For emission at the fundamental of the plasma frequency, we have two possible processes, analogous to those for Langmuir waves, namely the scattering by plasma ions, and the decay involving ion-sound waves. Ion-scattering was proposed in the original version of the plasma emission mechanism by \citet{GinzburgZhelezniakov1958}. The three-wave processes are generally faster, but if the ion-sound wave damping is very strong, they are suppressed. For kHz wavelengths, i.e. plasma emission in the solar wind, often the ion-temperature is low, so the three-wave processes are fast, and as shown by \citet{1993ApJ...416..831T} ion-scattering is too slow to explain the observed brightness temperatures. It is however sufficient in the corona, as shown by \citet{2003SoPh..215..335M}. 

The processes are the scattering of a Langmuir wave into an electromagnetic wave, $L + i \rightleftarrows t + i'$, and two processes involving ion-sound wave interactions, namely $L + s \rightleftarrows t$, $L \rightleftarrows t + s$. The latter was proposed by e.g.  
\citet{1986ApJ...308..954L} as the most likely generating process for ion-sound turbulence in solar wind, but it is now generally thought that Langmuir wave decay $L \rightleftarrows L' + s$ produces the ion-sound waves and these drive the production of fundamental electromagnetic emission \citep{Melrose1982,1994ApJ...422..870R}. 

The probability for ion-scattering which appears in Equation (\ref{eq:IS3D}) is \begin{equation}w_i^{LT}(\vec{k}_T, \vec{k}_L)= \frac{\sqrt{\pi} \omega_{pe}^2}{2 n_e v_{Ti} (1+T_e/T_i)^2}\frac{|\vec{k}_L \times \vec{k}_T|^2}{k_l^2 k_T^2|\vec{k}_T-\vec{k}_L|} \exp{\left(-\frac{(\omega_T-\omega_L)^2}{2|\vec{k}_T-\vec{k_L}|^2 v_{Ti}^2}\right)} \end{equation} 
identically to Equation (\ref{eq:cIon}). The main difference between this and the previous equation for Langmuir wave scattering is the cross product in the probability, which implies the emission peaks for scattering through 90$^o$.

For the three-wave scattering we have two alternatives
\begin{equation}\label{eqn:fundSprob} w^{LST}(\vec{k}, \vec{k}_L,\vec{k}_T)=\frac{\pi \omega_{pe}^3\left(1+\frac{3T_i}{T_e}\right)}{\omega_{k_T}^T 4n_eT_e}  \omega_k^S \frac{|\vec{k}_T\times\vec{k}|^2}{k_T^{2}|\vec{k}_L|^2}\delta(\omega_{k_T}^T-\omega^L_{k_L}\mp\omega^S_k)\end{equation} where the minus sign in the delta function is used for the process $L + s \rightleftarrows t$, the plus sign for $L \rightleftarrows t + s$. The only other difference between these two processes is a slight change in the participating wavenumbers, as we must consider $\vec{k}_t = \vec{k_L} \pm \vec{k}_s$ respectively. The three-wave probability also contains a cross product term, and is thus maximum for scattering by 90$^o$. In both cases (three-wave decay and ion-scattering) we have the condition $k_T \ll k_L$. 

If we consider a realistic situation in the solar corona or wind, we have a beam of electrons with some small pitch angle spread. The Langmuir waves this produces also have a small angular spread, and this is only increased by the presence of plasma density fluctuations \citep[e.g.][]{NishikawaRyutov1976}, and by the ion-scattering and three-wave decays acting on the waves. The fundamental emission can therefore be rather non-directional when it is emitted. However, from the expression for the group velocity, Equation (\ref{eq:vg}) and the dispersion relation, Equation (\ref{eq:dispL}), it is evident that emission close to the plasma frequency cannot propagate into plasma of increasing density. Ray tracing analyses by \citet{Li_etal2006c} showed, using typical type III source parameters that fundamental emission emitted into the forwards hemisphere at any angle  is rapidly redirected to propagate along the magnetic field direction.

\section{Conclusions} \label{sec:conclusion}

It is our opinion that solar radio physics is beginning to enter a new era.  Regarding type IIIs, new technology is allowing simulations to generate synthetic dynamic spectra for comparison to observations to probe relevant electron beam and background plasma properties.   At the same time new technology is giving birth to the next generation of radio telescope.  Interferometers with large baselines and huge collecting areas are being trained on the Sun o produce images of radio bursts over many frequencies with impressive angular resolution.  We conclude this article by summarising the future of type III radio burst analysis, both numerically and observationally.

\subsection{``State of the Art'' simulations}

While initially plasma emission work was mainly theoretical, simulations have always played a key role. Early work by e.g. \citet{TakakuraShibahashi1976,Takakura1982} allowed the simulation of bursts at a few wavelengths, and was key to the adoption of the ion-sound wave dependent model. 

Only recently have large-scale kinetic simulations become possible, as computing power is a strong limitation. The series of papers by \citet{2008JGRA..11306104L,2008JGRA..11306105L,2009JGRA..11402104L} were the first to trace an electron beam from the injection site into the corona and solar wind, and calculate the resulting radio emission fully numerically, rather than by analytical estimates. These simulations have more recently been extended to cover a wider frequency range, and used to explore the effects of the background plasma on the emission. In the solar wind, observations often show a kappa-distribution rather than a Maxwellian background plasma, and this contains more electrons of higher velocity. A similar power-law electron injection forms a higher velocity beam due to time-of-flight effects, and the emission may be expected to show a faster drift. This is confirmed by the simulations of \citet{LiCairns2014}, although these consider only instantaneous electron injection. 

Ongoing work by the authors aims to combine the large-scale simulations of e.g. \citet{ReidKontar2013} with a model for plasma radio emission used in a different context in \citet{2014A&A_RatcliffeKontar}. This aims to explore the effects of plasma density fluctuations on the emission. 

On the other hand, the fully numerical approach is not the only possibility. The simulations of \citet{RobinsonCairns1998,CairnsRobinson1999,Li_etal2006b,Li_etal2006c} used a combination of numerics and analytical estimates from Stochastic Growth Theory to reproduce emission. The relevance of SGT remains in question, but the results are useful nonetheless. The localisation of Langmuir waves into discrete clumps, or wave packets, is investigated using the Zakharov equations in e.g. \citet{Zaslavsky_etal2010,Krafft_etal2013}. 

Continued work on the propagation of electrons and their Langmuir wave generation and non-linear evolution is also essential. Langmuir waves can help to explain the observed non-Maxwellian thermal electron distribution in the solar wind \citep[e.g.][]{2012SSRv..173..459Y}. They may affect the hard-X ray emission from downward electron beams in the corona \citep[e.g.][]{Hannah_etal2013}. Plasma emission has even been suggested as a source of sub-THz radio emission \citep{2013AstL...39..650Z}.

\subsection{Next generation radio observations}

There are many solar telescopes all around the world that are able to observe the dynamic spectra of type III radio bursts, too many to mention individually.  On the global scale we would like to mention the e-Callisto project \citep{Benz_etal2009b}, an international network of solar radio spectrometers that has more than 66 instruments in more than 35 locations with users from more than 92 countries.

Regarding imaging, the most notable dedicated solar interferometer for type III radio bursts is the Nan\c{c}ay Radioheliograph \citep{KerdraonDelouis1997} (NRH), based in France, that has been making dedicated solar images since the 1960s and not images between 150 MHz and 450 MHz

The next generation radio observatories being developed around the world are taking observational solar radio astronomy into a revolutionary new phase.  Large baseline interferometers are making high resolution imaging spectroscopy observations of type III bursts.  This will enable us to address fundamental questions about the energy release site in solar flares and the transport of energetic electrons through the heliosphere.  In Europe there is the LOw Frequency ARray  \citep{vanHaarlem_etal2013} (LOFAR), a network of observatories spread across Europe (Netherlands, UK, Germany, France and Sweden).  LOFAR operates between the frequencies of 10 MHz and 250 MHz and is providing interferometric imaging with 10s arcsec resolutions.  In North America there is the recently upgraded Expanded Very Large Array \citep{Perley_etal2011} (EVLA) that operates between 1 GHz and 50 GHz and has started solar observing at the end of 2011.  In Western Australia there is the Murchison Widefield Array (MWA) \citep{Lonsdale_etal2009,Bowman_etal2013} that is a low-frequency radio telescope operating between 80 MHz and 300 MHz, is one of the precursors to the Square Kilometer Array (SKA) and has recently started to observe the Sun \citep{Oberoi_etal2011}.

Other next generation radio telescopes are on the horizon and will soon be ready for solar observations.  We now cover them (in no particular order).  In China the Chinese Spectral Radio Heliograph (CSRH) \citep{Yan_etal2009} will be operational in 2014 and will produce dedicated solar imaging between 400 MHz and 15 GHz, an ideal frequency range for type III bursts in the deep corona.  In Russia there is the Siberian Solar Radio Telescope (SSRT) that is being upgraded to create interferometric images of the Sun at GHz frequencies.  In India an upgrade is currently under way to the GMRT Giant Metrewave Radio Telescope (GMRT) to operate between the frequencies between 50 MHz and 1500 MHz.  In America, improvements to the existing array will create the Expanded Owens Valley Solar Array (EOVSA) that will operate between the frequency range 1 GHz to 18 GHz.  In South America we will have the Brazilian Decimeric Array that will operate between the frequencies of 1 GHz and 6 GHz.  In New Mexico there is the Long Wavelength Array (LWA) \citep{Lazio_etal2010,Taylor_etal2012} that will operate at the low frequencies between 10 - 88 MHz.

Extending into space, the upcoming ESA Solar Orbiter mission and NASA Solar Probe Plus mission are journeying close to the Sun to 10s solar radii, and should launch in 2017 and 2018 respectively.  Both missions will be armed with in-situ plasma measuring devices to explore the radial dependence of type III radio signals and the plasma properties of electron beams and the ambient solar wind.

\begin{acknowledgements}

This work is supported by a SUPA Advanced Fellowship (Hamish Reid), the European Research Council under the SeismoSun Research Project No. 321141 (Heather Ratcliffe), and the Marie Curie PIRSESGA- 2011-295272 RadioSun project.  

\end{acknowledgements}

\bibliographystyle{raa}
\bibliography{t3_reid_ratcliffe_ms1742}

\end{document}